\documentclass[pdflatex,sn-mathphys-num]{sn-jnl}

\usepackage{graphicx}%
\usepackage{multirow}%
\usepackage{amsmath,amssymb,amsfonts}%
\usepackage{amsthm}%
\usepackage{mathrsfs}%
\usepackage[mathscr]{euscript}
\usepackage[title]{appendix}%
\usepackage{xcolor}%
\usepackage{textcomp}%
\usepackage{manyfoot}%
\usepackage{booktabs}%
\usepackage{algorithm}
\usepackage[noend]{algpseudocode}
\usepackage{comment}
\usepackage{listings}%
\usepackage[justification=justified]{caption}
\usepackage[normalem]{ulem} 

\theoremstyle{thmstyleone}%
\newtheorem{theorem}{Theorem}
\newtheorem{lemma}{Lemma}
\theoremstyle{thmstyletwo}%

\theoremstyle{thmstylethree}%
\newtheorem{definition}{Definition}%

\usepackage{booktabs,array}
\usepackage{float}

\newcommand{\df}[1]{\textit{\textbf{#1}}}

\algnewcommand\algorithmicforeach{\textbf{foreach}}
\algdef{S}[FOR]{ForEach}[1]{\algorithmicforeach\ #1\ \algorithmicdo}

\usepackage{tikz}
\usetikzlibrary{arrows.meta, positioning}

\begin{document}

\title[Efficient Algorithms to Compute Closed Substrings]{Efficient Algorithms to Compute Closed Substrings}


\author[1]{\fnm{Samkith} \sur{K Jain}}\email{kishors@mcmaster.ca}
\author*[1]{\fnm{Neerja} \sur{Mhaskar}}\email{pophlin@mcmaster.ca}

\affil[1]{\orgdiv{Department of Computing and Software}, \orgname{McMaster University}, \orgaddress{\street{1280 Main St W}, \city{Hamilton}, \postcode{L8S 4L8}, \state{Ontario}, \country{Canada}}}

\abstract{A closed string $u$ is either of length one or contains a non-empty border that occurs only as a prefix and as a suffix in $u$ and nowhere else within $u$. In this paper, we present fast $\mathcal{O}(n\log n)$ time algorithms to compute all $\mathcal{O}(n^2)$ closed substrings by introducing a compact representation for all closed substrings of a string $ w[1..n]$, using only $\mathcal{O}(n \log n)$ space. These simple and space-efficient algorithms also compute maximal closed strings. Furthermore, we compare the performance of these algorithms and identify classes of strings where each performs best. Finally, we show that the exact number of maximal closed substrings in a Fibonacci word $ f_n $, for $n \geq 5$, is $\approx \left(1 + \frac{1}{\phi^2}\right) F_n \approx 1.382 F_n$, where $ \phi $ is the golden ratio and $F_n$ = $|f_n|$.}




\maketitle

\section{Introduction}
\label{sec:introduction}
A closed string $u$ is either of length one or contains a non-empty border that occurs only as a prefix and as a suffix in $u$ and nowhere else within $u$; otherwise, it is called open. In other words, the longest (possibly overlapping) border of a closed substring $u$ does not have internal occurrences in $u$. For example, in the string $abcab$, $ab$ occurs only as a prefix and a suffix. A maximal closed substring (MCS) is a closed substring that cannot be extended by one character to either the left or the right so as to form another closed substring. For example, in string $w[1..8] = abaccaba$, the MCSs are $w[1..8]=abaccaba$, $w[3..6]=acca$, $w[1..3]=w[6..8]=aba$, $w[4..5]=cc$ (five in total) and all occurrences of $a$ and $b$, while the substrings $w[2..7]=baccab$, $w[2..8]=baccaba$, $w[1..7]=abaccab$, and both occurrences of $c$ are closed but not maximal.

The concept of open and closed strings was introduced by Fici in~\cite{fici_2011_intro} to study the combinatorial properties of strings, with connections to Sturmian and Trapezoidal words. Then concepts of \textit{Longest Closed Factorization} and \textit{Longest Closed Factor Array} were introduced in~\cite{Badkobeh2014ClosedFactorization}, with improved solutions later proposed in~\cite{Bannai2015SPIRE}. Later,~\cite{Badkobeh2016DAM} introduced efficient algorithms for minimal and shortest closed factors. Related problems such as the $k$-closed string problem~\cite{Alamro2017EANN} and combinatorial problems on closed factors in Arnoux-Rauzy and $m$-bonacci words~\cite{JAHANNIA202232,PARSHINA201922} have also been explored. 

It has been shown that a string contains at most $\mathcal{O}(n^2)$ \textit{distinct} closed factors~\cite{Badkobeh2015LATA}. More recently (in 2024),~\cite{PARSHINA2024114315} showed that the maximal number of distinct closed factors in a finite word of length $n$ is at most $\frac{n^2}{6}$. In early 2025,~\cite{Mieno2025SOFSEM} presented a linear algorithm to count the number of distinct closed factors of a string in $\mathcal{O}(n \log \sigma)$ time and another in $\mathcal{O}(n)$ time. They also described methods to enumerate all closed and distinct closed factors, achieving a complexity of $\mathcal{O}(n^2)$, which is further optimized to $\mathcal{O}(n\sqrt{\log n} + \text{output} \cdot \log n)$ for distinct closed substrings, using weighted ancestor queries, where $\text{output}$ is the number of distinct closed substrings.

In~\cite{Badkobeh_2022_SPIRE} Badkobeh et al. introduced the notion of MCSs and proposed a $ \mathcal{O}(n \sqrt{n})$ loose upper bound on the number of MCSs in a string of length $n$. They also proposed an algorithm using suffix trees to compute the MCSs for strings on a binary alphabet in $ \mathcal{O}( n \log n + m)$ time, where $m$ is the number of MCSs. Badkobeh et al. in~\cite{BADKOBEH2026115628} extended their algorithm to a general alphabet with a time complexity of $\mathcal{O}(n \log n )$, implicitly improving the bound on the total number of MCSs, this was later directly proved by Kosolobov in~\cite{kosolobov2024closedrepeats}.

In this paper, we propose solutions to the following problems:

\noindent\textbf{Compact Representation of All $\mathcal{O}(n^2)$ Closed Substrings: } Given a string $w[1 .. n]$ we define a compact representation of all $\mathcal{O}(n^2)$ closed substrings of $w$ using $\mathcal{O}(n \log n)$ space. 

\noindent\textbf{Compute All $\mathcal{O}(n^2)$ Closed Substrings:}  
Given a string $w[1..n]$, we compute all $\mathcal{O}(n^2)$ closed substrings of $w$ in a compact representation in $\mathcal{O}(n \log n)$ time and $\mathcal{O}(n)$ auxiliary space.

\noindent\textbf{Compute All MCSs:} Given a string $w[1 .. n]$ we find all MCSs of $w$ in $\mathcal{O}(n \log n)$ time and $\mathcal{O}(n)$ auxiliary space.

\noindent\textbf{Exact Number of MCSs in a Fibonacci Word:} Given a Fibonacci word $f_n$, we give an exact equation to find the number of MCSs in $f_n$ which is denoted by $M(f_n)$ and is defined in Equation~\eqref{eq:MCS_count}. We also show that $M(f_n) \approx 1.382 F_n$.

In Section~\ref{sec:preliminaries}, we introduce the terminology and data structures used in the paper. In Section~\ref{sec:all_closed_strings} we define a compact representation of all closed substrings of a given string $w[1..n]$. In Section~\ref{sec:right_closed_ds_and_algo}, we present two algorithms to compute all closed substrings, including all MCSs. In Section~\ref{sec:enomifs} we provide a formula to compute the exact number of MCSs in a Fibonacci word. We then present experimental analysis of the proposed algorithms in Section~\ref{sec:experimental_eval}. Finally, we conclude with future work and open problems in Section~\ref{sec:conclusion_and_open_problems}.

An earlier and abridged form of this work appeared in \textit{SPIRE 2025}~\cite{KJAIN_SPIRE_2025}. In this extended version, we present an alternate algorithm to compute all closed substrings, including all MCSs, analyze the space complexities of both algorithms, and provide extensive experimental results comparing their performance across different classes of strings.

\section{Preliminaries}
\label{sec:preliminaries}
An alphabet $\Sigma$, of size $\sigma$, is a set of symbols that is totally ordered. A string (\df{word}), written as $w[1 .. n]$, is an element of $\Sigma^*$, whose length is represented by $n$. The \df{empty string} of length zero is denoted by $\varepsilon$. A \df{substring} (\df{factor}) $w[i .. j]$ of $w$ is a string, where $1 \leq i, j \leq n$, if $i > j$ then the substring is $\varepsilon$. A substring $w[i .. j]$ where $i = 1$ ($j=n$) is called a \df{prefix} (\df{suffix}) of $w$; it is a \df{proper prefix} (\df{suffix}) of $w$, if $j-i+1 < n$. If a substring $u$ of $w$ is both a proper prefix and a proper suffix of $w$, then it is called a \df{border}. A \df{cover} of a string $w$ is a substring $u$ of $w$ such that $w$ can be constructed by overlapping and/or adjacent instances of $u$. If $|u| < |w|$, then it is said to be a \df{proper cover} (for more details on covers, see the survey~\cite{NEERJA_STRING_COVERING}).

We now formally define the notions of closed strings and maximal closed strings introduced earlier. Let $k = \max\{ i \mid 0 \le i < n,\; w[1..i] = w[n-i+1..n]\}$ be the maximum border length of $w$. Then $w$ is \df{closed} if and only if $n = 1$ or $(k \ge 1 \text{ and } \forall j \in [2..n-k],\; w[1..k] \ne w[j..j+k-1])$. Otherwise, $w$ is \df{open}. An occurrence $w[i..j]$ of a substring $u$ in $w[1..n]$ is \df{maximal right-closed} if $u$ is closed and it cannot be extended to the right into another closed substring; that is, either $j = n$ or $w[i..j+1]$ is open. Symmetrically, the occurrence is \df{maximal left-closed} if $u$ is closed and either $i = 1$ or $w[i-1..j]$ is open. An occurrence $w[i..j]$ is a \df{maximal closed substring (MCS)} of $w$ if it is both maximal left-closed and maximal right-closed; that is, $u$ is closed, and $(i = 1 \text{ or } w[i-1..j] \text{ is open})$ and $(j = n \text{ or } w[i..j+1] \text{ is open})$. Hence, every MCS is a maximal right-closed substring, but the converse does not hold, since a maximal right-closed substring need not be maximal left-closed. An MCS of length one is called a \df{singleton} MCS; otherwise, it is called a \df{non-singleton} MCS. For example, in the string $w[1..8] = abaccaba$, the non-singleton MCSs are $w[1..8] = abaccaba$, $w[3..6] = acca$, $w[1..3] = w[6..8] = aba$, and $w[4..5] = cc$, while the singleton MCSs are all occurrences of $a$ and $b$.

An integer $p$ with $1 \le p \le n$ is called a \df{period} of $w[1..n]$ if $w[i] = w[i+p]$, for all $1\leq i \leq n-p$. We denote the \df{smallest period} of $w$ as $per(w)$. A string $w$ is \df{periodic} if $per(w) \le |w|/2$. A substring $w[i..j]$ is a \df{repetition} in $w$, if it is of the form $u^{e}$ for some non-empty string $u$, and integer $e \ge 2$; otherwise it is \df{primitive}. A substring $w[i..j]$ in a string $w$ is a \df{maximal repetition} if $w[i..j]$ is a repetition of the form $u^e$, $u$ is primitive ($|u|=t = per(w[i..j])$), and ($i\le t$ or $w[i-t..i+t(e-1)-1] \neq u^e$) and ($j+t > n$ or $w[i+t..i+(e+1)t-1] \neq u^e$). A \df{run} is a periodic substring $w[i..j] = u^{e} u'$ of $w[1..n]$, where $u$ is primitive ($|u|=t = per(w[i..j])$), $u'$ is a (possibly empty) proper prefix of $u$ and $e\geq 2$, whose periodicity cannot be extended to the left or to the right; that is, $i = 1$ or $w[i-1] \ne w[i+t-1]$, and $j = n$ or $w[j+1] \ne w[j+1-t]$.

A \df{gapped repeat} is a substring of $w[1..n]$ of the form $uvu$, where the two occurrences of $u$ are non-empty and are called the \df{arms}, $v$ is called the \df{gap}, and the number $|u| + |v|$ is the period. An occurrence of a gapped repeat is called a \df{maximal gapped repeat} if it cannot be extended to the left or to the right while preserving its period. A maximal gapped repeat that is closed is referred to as a \df{gapped-MCS}.

In the literature, a gapped repeat is formally a pair consisting of a substring together with its period $|u|+|v|$, since the same substring may admit several decompositions $uvu$ with different periods. Instead, we refer to a gapped repeat simply as a substring of the form $uvu$. This convention differs slightly from the standard one, but suffices for our purposes.

A closed substring $u = w[i..j]$ of a string $w[1..n]$ is said to be \df{right-extendible} to a maximal right-closed substring $r = w[i..j']$ if $j < j' \leq n$ and every prefix $w[i..k]$ of $r$ with $j < k < j'$ is also closed. In this case, $r$ is the unique maximal right-closed extension of $u$.

The \df{suffix array $\mathsf{SA}[1 .. n]$} of $w[1..n]$ is an integer array of length $n$, where each entry $\mathsf{SA}[i]$ points to the starting position of the $i$-th lexicographically least suffix of $w$. The \df{longest common prefix array $\mathsf{LCP}[1 .. n]$} of $w[1..n]$ is an array of integers such that $\mathsf{LCP}[i]$ is equal to the length of the longest common prefix of suffixes starting at positions $\mathsf{SA}[i-1]$ and $\mathsf{SA}[i]$ for all $2 \leq i \leq n$, we assume $\mathsf{LCP}[1] = 0$. 

\subsection{Overview of Crochemore’s Maximal Repetition Algorithm}\label{ssec:overviewCMR}

We provide a brief overview of Crochemore's well-known algorithm for computing maximal repetitions of a given string $w$~\cite{CROCHEMORE1981244}--which we refer to as \df{CMR Algorithm}, while highlighting key facts that form the basis of the algorithm. This algorithm serves as the basis for our Algorithm~\ref{alg:mrc_crochemore}.

\noindent\textit{Fact 1 (Levels):} The \df{CMR} algorithm operates on levels $k = 1, 2, \ldots, n$, where $|w|=n$. At each level $k$, it maintains a collection of disjoint sets $C_k$ of index positions corresponding to starting positions  of substrings of $w$ of length $k$, grouped according to substring equality. 

Let $1 \leq k \leq n$ denote a level in the CMR Algorithm. An equivalence relation $ \approx_k $ on the set of positions $ \{1, 2, \ldots, n\} $ is defined by 
\[
i \approx_k j \Longleftrightarrow w[i..i+k-1] = w[j..j+k-1],
\]
where $ i, j \in \{1, 2, \ldots, n\}$ are distinct index positions. The corresponding equivalence classes of $ \approx_k $ form a collection of disjoint sets whose union is $\subseteq \{1,2, \ldots n\}$. For each position $ i $, we denote its equivalence class by 
\[
[i]_k = \{\, j \mid j \approx_k i \,\}
\]
Thus, each class $ [i]_k $ consists of the starting positions of all occurrences of a substring of length $ k $ in $ w $. Let $U_k$ be the set of all distinct substrings of $w$ of length $k$. Then, the set of equivalence classes at level $k$ is defined as 
\[
C_k = \Bigl\{ \{ j \mid w[j..j+k-1] = u \} \Bigm| u \in U_k \Bigr\}.
\]
Figure~\ref{fig:eClasses} shows the classes of $\approx_k$ where $ 1 \leq k \leq 5$ for the string $w[1..11] = mississippi$.

\noindent\textit{Fact 2 (Class Representation).} 
For each $k = 1, 2, \ldots, n$, all sets in $C_k$ are stored as ordered lists. Consequently, in any class $C' \in C_k$, the index positions appear in sorted order. The ordered list representation is implemented using doubly linked lists with auxiliary pointers to elements, along with additional auxiliary data structures, allowing constant-time insertion, deletion, and next and previous element lookup. 

\noindent\textit{Fact 3 (Refinement):} The equivalence classes at level $k+1$, for $1\le k\le n-1$, are computed from those at level $k$ through a process called \df{refinement}.

Initially, the equivalence classes at level $k = 1$ are obtained by scanning the string $w$ from left to right and grouping together positions that contain the same symbol in $\Sigma$. Therefore, at $k=1$, we obtain $\sigma$ classes. To compute the classes at level $k+1$, each class in $C_k$ is \textit{refined} as follows: let $C',C'' \in C_k$ such that $C' \ne C''$. For a class $C'$, consider indices $i, j \in C'$ such that $i \ne j$. If $i+1, j+1 \in C''$, then $i$ and $j$ remain in the same class; otherwise they are separated or \textit{refined out} into different possibly new classes. 

Once we refine all classes at level $k$, we get all the classes at level $k+1$. The refinement process to compute equivalence classes at various levels for the string $w = mississippi$ is illustrated with arrows in Figure~\ref{fig:eClasses}. 

\noindent\textit{Fact 4 (Termination):} Once a class becomes a singleton (of size one), it cannot be refined further. The algorithm terminates when all classes are singletons.

The $\mathcal{O}(n \log n)$ time complexity, in Lemma~\ref{lem:crochemore_complexity}, lies in the choice of classes for refinement. For each family of classes (classes that were formed as a refinement of a class on the previous level), we identify the largest class and call all the others \df{small}. By using only small classes for refinement to get all the equivalence classes in the next level, $\mathcal{O}(n \log n)$ complexity is achieved, as each element belongs only to $\mathcal{O}(\log n)$ small classes. To support efficient refinement operations, the classes are maintained as ordered doubly linked lists, allowing elements to be removed from one class and inserted into another in constant time. For a detailed description of the CMR algorithm and its complexity, we refer the reader to~\cite{CROCHEMORE1981244}.

\begin{lemma}[Theorem 7 in \cite{CROCHEMORE1981244}]
\label{lem:crochemore_complexity}
The CMR Algorithm correctly computes all the maximal repetitions in a string $w[1..n]$ in $\mathcal{O}(n \log n)$ time.
\end{lemma}

\section{Compact Representation of all Closed Substrings}
\label{sec:all_closed_strings}
A string $w[1 .. n]$ contains $\mathcal{O}(n^2)$ closed substrings, which can naively be computed in $\mathcal{O}(n^2)$ time by computing the border of all substrings of $w$. In this section, we define a compact representation for all $\mathcal{O}(n^2)$ closed substrings in Theorem~\ref{thm:compact_representation_and_space_bound} that requires only $\mathcal{O}(n \log n)$ space. Algorithm~\ref{alg:compact_representation} in Section~\ref{sec:right_closed_ds_and_algo} presents an $\mathcal{O}(n \log n)$ time solution to compute all closed substrings of $w$ in a compact representation of size $\mathcal{O}(n \log n)$.

Let $u$ and $v$ be proper closed prefixes of the maximal right-closed substring $r$. Then, by definition of right-extendibility we define an equivalence relation $R$ as follows: $$R = \{(u, v) \mid u \text{ and } v \text{ are both right-extendible to } r\}.$$ Next, we define the equivalence class $E_{r}$ of $r$ under the relation $R$ as follows:
\begin{equation}
\label{eq:equivalence_class} 
E_r = \{ u \mid u \text{ is right-extendible to } r \}.
\end{equation}
Note that by definition of right-extendibility $r$ does not belong to $E_{r}$.
\begin{lemma}
\label{lem:shortest_right_closed_string_new}
Given a string $w[1 .. n]$, if $w[1..m]$ is the shortest maximal right-closed prefix of $w$, then $w[1..m] = \lambda^m $, where $ \lambda \in \Sigma $ and $ 1 \leq m \leq n$. Moreover, each prefix of $ \lambda^m $ is also closed. 
\end{lemma}

\begin{proof}
We prove the lemma in two parts. 
First part by contradiction. Suppose that $w[1..m]$ is the shortest maximal right-closed prefix of $w$, and that $w[1 .. m] \neq \lambda^m$. W.l.o.g let $ w[1] = \lambda$. Let $1 < j \leq m$ be the smallest index such that $ w[j] = \lambda_1$ and $w[1..j-1] = \lambda^{j-1}$, where $\lambda_1 \neq \lambda \in \Sigma$. Then, $w[1..j] = \lambda^j \lambda_1 $. However, by definition of maximal right-closed string, $w[1..j-1] = \lambda^{j-1}$ is a shorter maximal right-closed string of $w$ --- a contradiction. 

Second, for any string of the form $ u = \lambda^m $, consider its prefixes $ p = u[1 .. j] $, where $ 1 \leq j \leq m $. If $ j = 1 $, then $ p $ is closed trivially. If $ j \geq 2 $, then the border length of each prefix $p$ is equal to $j-1$, which is maximum and unique. Thus, all the prefixes of $ u = \lambda^m $ are closed. 
\end{proof}

\textit{Let $ r $ denote a maximal right-closed substring, $ b $ its border, and $ p $ its shortest closed prefix such that $p = r$ or $ p \in E_r $.}
Let $ r_1, r_2, \ldots, r_{K_i} $ be all the maximal right-closed substrings starting at index $ i $ of a string $ w[1 .. n] $, where $ 1 \leq i \leq n $, and $ K_i $ is the number of maximal right-closed substrings starting at index $ i $, such that $ |r_1| < |r_2| < \cdots < |r_{K_i}| $. 

\begin{figure}[t]
\centering
\begin{tikzpicture}
\small
\draw (-6,-0.25) rectangle (6, 0.25);

\draw (-6,-0.25) rectangle (-4.5,0.25); 
\draw (-6,-0.25) rectangle (-3,0.25);
\draw (-3,-0.25) rectangle (-1.5,0.25);
\draw (0,-0.25) rectangle (1.5,0.25);
\draw (1.5,-0.25) rectangle (2,0.25);

\draw[<->] (-6, -2.25) -- (6, -2.25) node[midway, above, yshift=-2pt] {$r_j = p_jp'_j$};
\draw[<->] (-6, -0.75) -- (-1.5, -0.75) node[midway, above, yshift=-2pt] {$r_{j-1}$};
\draw[<->] (0, -0.75) -- (2, -0.75) node[midway, above, yshift=-2pt] {$|b_{j-1}| + 1$};
\draw[<->] (-6, -1.75) -- (2, -1.75) node[midway, above, yshift=-2pt] {$p_j$};
\draw[<->] (2, -1.75) -- (6, -1.75) node[midway, above, yshift=-2pt] {$p'_j$};
\draw[<->] (0, -1.25) -- (6, -1.25) node[midway, above, yshift=-2pt] {$b_j$};

\draw[dashed] (-6,-0.25) -- (-6,-2.25);
\draw[dashed] (6,-0.25) -- (6,-2.25);
\draw[dashed] (-1.5,-0.25) -- (-1.5,-0.75);
\draw[dashed] (2,-0.25) -- (2,-1.75);
\draw[dashed] (0,-0.25) -- (0,-1.25);

\draw node at (-3.75,0) {$\dots$};
\draw node at (-0.75,0) {$\dots$};
\draw node at (-5.25,0) {$b_{j-1}$};
\draw node at (-2.25,0) {$b_{j-1}$};
\draw node at (0.75,0) {$b_{j-1}$};

\draw node at (-6,0.5) {$i$};

\end{tikzpicture}
\caption{Illustration of the second case in the proof of Lemma~\ref{lem:each_prefix_is_closed}, showing that $|p_j|$ can be computed from $|b_j|$, $|b_{j-1}|$ and $|r_j|$, such that $p_j = r_j$ or $p_j \in E_{r_j}$. Note that when $p_j = r_j$, $p_j' = \varepsilon$.}
\label{fig:each_prefix_is_closed_illustration}
\end{figure}

In Equation~\eqref{eq:length_of_prefix} we present a formula to compute the length of the shortest closed prefix $p_j$ of $r_j$ such that $p_j = r_j$ or $p_j \in E_{r_j}$, where $1 \leq j \leq K_i$. Lemma~\ref{lem:each_prefix_is_closed} proves Equation~\eqref{eq:length_of_prefix} and is illustrated in Figure~\ref{fig:each_prefix_is_closed_illustration}.

\begin{lemma}
\label{lem:each_prefix_is_closed}
Let $1 \leq j \leq K_i$. The length of the shortest closed prefix $p_j$ of $r_j$, such that $p_j = r_j$ or $p_j \in E_{r_j}$, is given by:
\begin{equation}
|p_j| =
\begin{cases} 
1, & \text{if } j = 1, \\
|r_j| - |b_j| + |b_{j-1}| + 1, & \text{if } 1 < j \leq K_i,
\end{cases}
\label{eq:length_of_prefix}
\end{equation}
where $b_j$ and $b_{j-1}$ are the longest borders of $r_j$ and $r_{j-1}$, respectively.
\end{lemma}

\begin{proof}
We prove the lemma by considering two cases. For $j = 1$, by Equation~\eqref{eq:equivalence_class} and Lemma~\ref{lem:shortest_right_closed_string_new}, all prefixes of $r_1$ are closed and $|p_1| = 1$.

For $1 < j \leq K_i$, by definition of a closed substring, the shortest closed prefix $p_j$ starting at index $i$ and longer than $r_{j-1}$ (whose longest border has length $|b_{j-1}|$) must have the longest border of length exactly $|b_{j-1}| + 1$. Either $p_j = r_j$ or by definition of right-extendibility $p_j$ extends to $r_j = p_j p_j'$, with the longest border $b_j$, and each prefix of $r_j$ with length at least $|p_j|$ is closed. Since $|p_j'| = |b_j| - (|b_{j-1}| + 1)$, we have $|r_j| = |p_j| + |p_j'| = |p_j| + |b_j| - (|b_{j-1}| + 1)$. Solving for $|p_j|$, we obtain the desired formula: $|p_j| = |r_j| - |b_j| + |b_{j-1}| + 1$. 
\end{proof}

\begin{lemma}[Theorem~1 in~\cite{kosolobov2024closedrepeats}]
\label{lem:bound_on_no_maximal_lr_substrings}
A string $w[1..n]$ contains at most $\mathcal{O}(n \log n)$ maximal right (left)-closed substrings.
\end{lemma}

Theorem~\ref{thm:compact_representation_and_space_bound} follows directly from Equation~\eqref{eq:equivalence_class}, Lemma~\ref{lem:each_prefix_is_closed} and~\ref{lem:bound_on_no_maximal_lr_substrings}.

\begin{theorem}
\label{thm:compact_representation_and_space_bound}
The compact representation of all closed substrings of $w[1..n]$ requires at most $\mathcal{O}(n \log n)$ space and is given by:
\begin{equation}
\label{eq:compact_representation}
\mathcal{C}(w) = \{ (i, |p_j|, |r_j|) \mid 1 \leq i \leq n,\ 1 \leq j \leq K_i \},
\end{equation}
where $K_i$ is the number of maximal right-closed substrings starting at position $i$, and for each $j$-th such substring $r_j$, starting at index $i$, $p_j$ is the shortest closed prefix of $r_j$ such that $p_j = r_j$ or $p_j \in E_{r_j}$.
\end{theorem}

The set of all closed substrings of $w[1..n]$, denoted by $\text{Closed}(w)$, can be enumerated using the following equation:
\begin{equation}
\label{eq:closed_string_enumeration}
\text{Closed}(w) = \bigcup_{(i, |p_j|, |r_j|) \in \mathcal{C}(w)} \left\{ w[i \ldots i + \ell - 1] \mid \ell \in [|p_j|, |r_j|] \right\}.
\end{equation}
An example of $\mathcal{C}(w)$ and the enumeration $\text{Closed}(w)$ for the string $w[1..11] = mississippi$ is given in Table~\ref{tab:compact_repr_mississippi}.

\section{Algorithms for Closed Substrings and MCSs}
\label{sec:right_closed_ds_and_algo}

In this section, we introduce the primary data structure--the $\mathcal{MRC}$ array--which is used to compute all closed substrings in a compact representation and MCSs in $\mathcal{O}(n \log n)$ time. See Table~\ref{tab:mrc_example} for an example of the $\mathcal{MRC}$ array for the string $w[1..11]=mississippi$. We present two algorithms: Algorithm~\ref{alg:mrc} based on $\mathsf{SA}$ \& $\mathsf{LCP}$ array and Algorithm~\ref{alg:mrc_crochemore} based on Crochmore's equivalence classes approach.

\begin{definition}[$\mathcal{MRC}$ Array\footnote{This definition uses the notation introduced in Section~\ref{sec:all_closed_strings}.}]
Given a string $w[1 .. n]$, the \df{maximal right-closed array ($\mathcal{MRC}$)} is an array of size $n$, where $\mathcal{MRC}[i]$, $1 \leq i \leq n$, stores the list of ordered pairs $(|r_j|, |b_j|)$, where $j=K_i$ to $1$.
\end{definition}

\newcolumntype{L}[1]{>{\hspace{0.8cm}\raggedright\arraybackslash}p{#1}}
\begin{table}[h]
\centering
\begin{tabular}{@{}c | L{3.5cm}@{}}
\toprule
\textbf{Index ($i$)} & \centering\arraybackslash \textbf{$\mathcal{MRC}[i]$} \\
\midrule
$1$  & $(1, 0)$ \\
$2$  & $(7, 4), (1, 0)$ \\
$3$  & $(6, 3), (2, 1)$ \\
$4$  & $(5, 2), (3, 1), (1, 0)$ \\
$5$  & $(4, 1), (1, 0)$ \\
$6$  & $(2, 1)$ \\
$7$  & $(1, 0)$ \\
$8$  & $(4, 1), (1, 0)$ \\
$9$  & $(2, 1)$ \\
$10$ & $(1, 0)$ \\
$11$ & $(1, 0)$ \\
\bottomrule
\end{tabular}
\caption{The $\mathcal{MRC}$ array for the string $w[1..11] = mississippi$.}
\label{tab:mrc_example}
\end{table}

\subsection{Computing \texorpdfstring{$\mathcal{MRC}$}{MRC} Array using \texorpdfstring{$\mathsf{SA}$}{SA} and \texorpdfstring{$\mathsf{LCP}$}{LCP}}\label{ssec:mrcsalcp}

Algorithm~\ref{alg:mrc} computes the $\mathcal{MRC}$ array by first identifying all the maximal right-closed substrings of length greater than one. It uses a stack to store AVL trees created during its execution. The algorithm scans the $\mathsf{LCP}$ array from left to right. Beginning with $\mathsf{LCP}[1]=0$, and for increasing $\mathsf{LCP}[i]$ values, the algorithm creates a single node AVL tree with $key=\mathsf{SA}[i]$ and $value=\mathsf{LCP}[i]$ and pushes it onto the stack, until either the $LCP$ value decreases or the end of the array is reached. 

We denote by $LCP_{max}$ the value stored in the root node of the AVL tree on the top of the stack. When a decrease in the $LCP$ value (or the end of the array) is encountered, the algorithm pops a collection of AVL trees from the stack, and store them in $\mathcal{L}_{\mathit{suffixSets}}$. Note that all trees in $\mathcal{L}_{\mathit{suffixSets}}$ have the following property: the suffixes represented by the keys in each AVL tree share a common prefix of length $LCP_{max}$. Conversely, any suffix of $w$ that shares a common prefix of length $LCP_{max}$ with the suffixes represented in $\mathcal{L}_{\mathit{suffixSets}}$ is represented by a node in one of the AVL trees in $\mathcal{L}_{\mathit{suffixSets}}$. Moreover, since $LCP_{max}$ is the maximum $LCP$ value among all the AVL trees on the stack, this collection stored in the list $\mathcal{L}_{\mathit{suffixSets}}$, preserves the strictly decreasing order of ordered pairs in the $\mathcal{MRC}$ array. This collection is then merged into a single AVL tree using the \textsc{Union} operation presented in~\cite{Jakub2015}\footnote{The \textsc{Union} operation, defined in~\cite{Jakub2015}, builds upon foundational methods from~\cite{brodal2000,tarjan1979}.}, resulting in $\mathcal{T}_{\mathit{{suffixes}}} = \bigcup \mathcal{L}_{\mathit{suffixSets}}$. 

We minimally modify the \textsc{Union} operation to also output a list of ordered pairs representing maximal right-closed substrings via the \textsc{ChangeList} operation defined in Equation~\ref{eq:change_list}. Let $\mathcal{P}' \subseteq [1..n]$, and let $\mathcal{P} = \{P_1, P_2, \ldots, P_t\}$ be a partition of $\mathcal{P}'$; that is, $\bigcup\limits_{i=1}^{t} P_i = \mathcal{P}'$, $P_i \cap P_j = \emptyset$ for $i \neq j$, and $P_i \neq \emptyset$ for all $i$. For $X \subseteq \mathbb{Z}^+$ and $x < \max(X)$, define $next_X(x) = \min\{ y \in X \mid y > x\}$. For $x = \max{(X)}$, $next_X(x)$ is not defined. We then define:

\begin{equation}\label{eq:change_list}
ChangeList(\mathcal{P},\mathcal{P}') = \{ (x, next_{\mathcal{P}'}(x)) \mid{}
  next_{P_i}(x) \neq next_{\mathcal{P}'}(x), x \in P_i\},
\end{equation} where $i$ is the unique index with $x\in P_i$.

By construction, the AVL trees stored in $\mathcal{L}_{\mathit{suffixSets}}$ is a partition of the set of suffix positions represented by $\mathcal{T}_{\mathit{suffixes}}$. Hence, $\mathcal{L}_{\mathit{suffixSets}}$ and $\mathcal{T}_{\mathit{suffixes}}$ correspond to the partition $\mathcal{P}$ and the set $\mathcal{P}'$, respectively, in the definition of \textsc{ChangeList}.
Each pair $(x, y) \in ChangeList(\mathcal{L}_{\mathit{suffixSets}}, \mathcal{T}_{\mathit{{suffixes}}})$ identifies a maximal right-closed substring $r_j=w[x..y + LCP_{max} - 1]$, where $LCP_{max}$ is the length of its longest border. Hence, the pair $(y + LCP_{max} - x ,LCP_{max})$ is inserted into $\mathcal{MRC}[x]$. Since $\mathcal{MRC}[x]$ is a list, the insert operation is performed in constant time. All such pairs from the $\textsc{ChangeList}$ are inserted into the $\mathcal{MRC}$ array. The algorithm then assigns to the root node of the AVL tree, $\mathcal{T}_{\mathit{{suffixes}}}$, the $LCP$ value of the AVL tree at the top of the stack and pushes it onto the stack, only if the stack is not empty--thus maintaining the lexicographic ordering of the set of suffixes--and continues scanning the $\mathsf{LCP}$ array. 

To compute the maximal right-closed substrings of length one, it performs a simple linear scan of the string $w$ and adds $w[i]$ to $\mathcal{MRC}[i]$, if $i=n \lor (1 \leq i < n \land w[i] \ne w[i+1])$.

It is known that $\mathsf{LCP}$ array identifies all the occurrences of the repeating substrings that cannot be extended to the right. Moreover, the suffix array groups all these occurrences within specific ranges. Algorithm~\ref{alg:mrc} identifies such ranges by tracking the rise and fall of values in the $\mathsf{LCP}$ array, and stores the corresponding indices in $\mathcal{L}_{\mathit{suffixSets}}$ as keys in AVL trees, starting with the substring of length $LCP_{max}$. The \textsc{Union} operation then merges these single node AVL trees to form one AVL tree ($\mathcal{T}_{\mathit{{suffixes}}}$) that contains all the $\mathsf{SA}$ values; that is, all the starting positions of the substrings of length $LCP_{max}$ in the string $w$.

The $\textsc{ChangeList}(\mathcal{L}_{\mathit{suffixSets}}, \mathcal{T}_{\mathit{{suffixes}}})$ operation then returns every pair of consecutive occurrences of substrings in the $\mathcal{L}_{\mathit{suffixSets}}$. These consecutive occurrences form the border of the maximal right-closed substring and are therefore added to the $\mathcal{MRC}$ array. Finally, a simple linear scan adds maximal right-closed substrings of length one, as shown in the \textsc{AddSingletonMRC} procedure presented in Algorithm~\ref{alg:addSingletons}. The algorithm clearly computes the maximal right-closed substrings in strictly decreasing order of length as required in the $\mathcal{MRC}$ array. Thus, the correctness of Algorithm~\ref{alg:mrc} follows.

\begin{algorithm}[t]
\caption{Add all maximal right-closed substrings of length $1$ to the $\mathcal{MRC}$ array.}
\label{alg:addSingletons}
\begin{algorithmic}[1]
\Procedure{AddSingletonMRC}{$w[1..n], \mathcal{MRC}$}
\For{$i \gets 1$ \textbf{to} $n$}
    \If{$i = n \;\textbf{or}\; w[i] \ne w[i+1]$}
        \State insert($\mathcal{MRC}[i]$, $(1, 0)$)
    \EndIf
\EndFor
\EndProcedure
\end{algorithmic}
\end{algorithm}

In Algorithm~\ref{alg:mrc}, the maximum size of the stack is at most $n$. Moreover, the total space occupied by all the AVL trees is also $n$, as the AVL trees form a disjoint partition of suffixes starting at index positions $1$ to $n$. Therefore, the maximum size of the structure $\mathcal{L}_{\mathit{\mathit{suffixSets}}}$ is bounded by $n$. The procedure $\textsc{ChangeList()}(\mathcal{L}_{\mathit{\mathit{suffixSets}}}, \mathcal{T}_{\mathit{\mathit{{suffixes}}}})$ can return a list of at most $n$ pairs because no substring of $w[1..n]$ can repeat more than $n$ times. Hence, Algorithm~\ref{alg:mrc} requires $\mathcal{O}(n)$ auxiliary space.

We note that the algorithm in the worst case performs $4n-2$ stack operations for a string $w=\lambda^n$ and in the best case performs $2n$ stack operations for a string with all distinct characters.
\begin{algorithm}[t]
\caption{Compute $\mathcal{MRC}$ Array of $w[1..n]$ using its $\mathsf{SA}$ and $\mathsf{LCP}$}
\label{alg:mrc}
\begin{algorithmic}[1]
\Procedure{Compute-MRCArray1}{$w, SA, LCP$}
\State $ i \gets 0 $; $\mathcal{MRC} \gets \{\emptyset\}^{n}$; $stack \gets \emptyset$
\While{$stack \neq \emptyset \textbf{ or } i < n $}
    \While{$ i < n \textbf{ and }(stack = \emptyset \textbf{ or } \mathsf{LCP}[i] \geq LCP(top(stack))$} 
        \State $push(stack, AVLTree(\{\mathsf{SA}[i]\}, \mathsf{LCP}[i]))$
        \State $ i \gets i + 1 $
    \EndWhile

    \State $\mathcal{L}_{\mathit{suffixSets}} \gets \emptyset$
    \State $LCP_{max} \gets LCP(top(stack))$
    \While{$stack \neq \emptyset  \textbf{ and } LCP(top(stack)) = LCP_{max}$}
        \State $insert(\mathcal{L}_{\mathit{suffixSets}}, pop(stack))$
    \EndWhile
    \If{$LCP_{max} \neq 0$}
        \State $previousLCP \gets LCP(top(stack))$
        \State $insert(\mathcal{L}_{\mathit{suffixSets}}, pop(stack))$
        \State $\mathcal{T}_{\mathit{{suffixes}}} = \bigcup \mathcal{L}_{\mathit{suffixSets}}$ \Comment{$\bigcup \mathcal{L}_{\mathit{suffixSets}}$ returns an AVL tree}
    
        \ForEach{$(x, y) \in \textsc{ChangeList}( \mathcal{L}_{\mathit{suffixSets}}, \mathcal{T}_{\mathit{{suffixes}}})$} \Comment{Equation~\eqref{eq:change_list}}
            \State $insert(\mathcal{MRC}[x], (y + LCP_{max} - x ,LCP_{max}))$
        \EndFor
    
        \State $ LCP(\mathcal{T}_{\mathit{{suffixes}}}) \gets previousLCP $
        \State $push(stack, \mathcal{T}_{\mathit{{suffixes}}})$
    \EndIf
\EndWhile
\State \textsc{AddSingletonMRC}($w$, $\mathcal{MRC}$) \Comment{See Algorithm~\ref{alg:addSingletons}}
\State \Return $\mathcal{MRC}$
\EndProcedure
\end{algorithmic}
\end{algorithm}

\begin{lemma}[\cite{Jakub2015}]
\label{lem:union_complexity}
The \textsc{Union} operation correctly computes the ChangeList, and any sequence of \textsc{Union} operations takes $\mathcal{O}(n \log n)$ time in total.
\end{lemma}

\begin{theorem}
\label{thm:mrc_complexity}
Algorithm~\ref{alg:mrc} correctly computes the $\mathcal{MRC}$ array of a string $w[1 .. n]$ using $\mathsf{SA}$ and $\mathsf{LCP}$ in $\mathcal{O}(n \log n)$ time and $\mathcal{O}(n)$ auxiliary space.
\end{theorem}

\begin{proof}
    We prove the total time complexity of Algorithm~\ref{alg:mrc} in three parts. First, for stack operations, we introduce a potential function $\Phi(S) = |S| \geq 0$, where $|S|$ is the number of elements in the stack. The actual cost of each push and pop operation is 1. Since we push all $n$ single node AVL trees corresponding to each suffix onto the stack, the amortized cost for each push operation is $c'_{push} = 1 + \Delta \Phi(S) = 2$, resulting in a total amortized cost of $2n$. Additionally, in the algorithm, we pop at least two AVL trees and push their \textsc{Union} back onto the stack, and because the amortized cost of this sequence of pops and push in the worse case is $c'_{pops-push} = 3 + \Delta \Phi(S) = 2$. The maximum number of such operations is $n - 1$, leading to a total amortized cost of $2(n - 1)$. Therefore, the total cost of all stack operations in the worst case is $4n - 2$, which is $\mathcal{O}(n)$. Second, by Lemma~\ref{lem:union_complexity}, the time complexity of the sequence of \textsc{Union} operations is $\mathcal{O}(n \log n)$. Finally, the single-characters that are maximal right-closed are computed by a linear scan on $w$, taking $\mathcal{O}(n)$ time. Therefore, the total time complexity is $\mathcal{O}(n \log n)$. 
\end{proof}

\subsection{Computing \texorpdfstring{$\mathcal{MRC}$}{MRC} Array using Equivalence Classes\texorpdfstring{~\cite{CROCHEMORE1981244}}{}}

Algorithm~\ref{alg:mrc} requires the pre-processed $\mathsf{SA}$ and $\mathsf{LCP}$ arrays to compute the $\mathcal{MRC}$ array of a string $w[1..n]$. In this section, we present an alternative algorithm that computes the $\mathcal{MRC}$ array directly from the input string $w[1..n]$, without any pre-processing, by modifying the CMR Algorithm that uses the notion of equivalence classes.

\begin{figure}[htbp!]
    \centering
    \includegraphics[width=\linewidth]{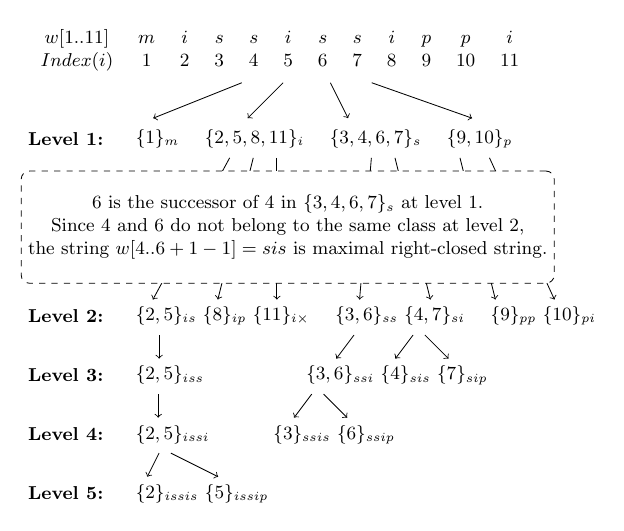}
    \caption{Equivalence classes across successive levels in the CMR Algorithm for the string $w[1..11] = mississippi$, illustrating the identification of maximal right-closed substrings from level transitions.}
    \label{fig:eClasses}
\end{figure}

The CMR Algorithm identifies the starting index positions of all unique substrings of $w[1..n]$ of length $k$, for increasing values of $k$ starting with $k=1$. The combined occurrences of substrings at consecutive index positions within a class $C' \in C_k$ then form a closed substring (not necessarily a maximal right-closed substring). We first describe how to identify all maximal right-closed substrings from the equivalence classes across all the $k$ levels in the CMR Algorithm, and then present constant-time modifications to it that enable computation of the $\mathcal{MRC}$ array in $\mathcal{O}(n \log n)$ time.

Recall from Section~\ref{ssec:mrcsalcp} that, $next_X(x) = \min\{ y \in X \mid y > x\}$, where $X \subseteq \mathbb{Z}^+$ and $x < \max(X)$. Similarly, we define $prev_X(x)=\max\{y \in X \mid y<x\},$ where $x>\min(X)$. For $x = \min{(X)}$, $prev_X(x)$ is not defined. Consider a class $C' \in C_k$. For an index $i \in C'$, the value $next_{C'}(i)$ (if defined) denotes the successor of $i$, that is, the smallest element greater than $i$ in the sorted class list $C'$. Let $j = next_{C'}(i)$. Equivalently, $i = prev_{C'}(j)$. The pair $(i,j)$ forms a closed substring starting at position $i$ with a border of length $k$. Note that both $next_{C'}()$ and $prev_{C'}()$ can be computed in constant time for any class $C' \in C_k$, as shown in~\cite{CROCHEMORE1981244}.

At the next level $k + 1$, suppose $i \in C''$ for some $C'' \in C_{k+1}$ but $j \notin C''$. This implies that the new characters appended to substrings of length $k$ at positions $i$ and $j$ are not the same, and therefore the border of length $k$ cannot be extended to the right. Hence, the substring starting at index position $i$ of length $j + k - i$ is a maximal right-closed substring. Conversely, suppose that $w[i..j+k-1]$ is a maximal right-closed substring whose longest border has length $k$. Then $i$ and $j$ belong to the same equivalence class $C' \in C_k$, and $j = next_{C'}(i).$ For example, in Figure~\ref{fig:eClasses}, which illustrates the equivalence classes for the string $w[1..11] = mississippi$, $6$ is the successor of $4$ in the class $\{3,4,6,7\}_s$ at level $1$. Since $4$ and $6$ do not belong to the same class at level $2$, the substring $w[4..6 + 1 - 1] = sis$ is a maximal right-closed substring with unique border $w[4..4] = [6..6] = s$ of length $k = 1$.

\begin{algorithm}[htbp!]
\caption{Compute $\mathcal{MRC}$ array of $w[1..n]$ without pre-processing $w$}
\label{alg:mrc_crochemore}
\begin{algorithmic}[1]
\Procedure{Compute-MRCArray2}{$w$}
\State $\mathcal{MRC} \gets \{\emptyset\}^{n}$
\State \textsc{AddSingletonMRC}($w$, $\mathcal{MRC}$) \Comment{See Algorithm~\ref{alg:addSingletons}}
\State $T[1..n] \gets \{0\}^n$; $k \gets 1$
\State $C_1 \gets$ Equivalence classes at level $1$, all marked \textit{small}.
\While{level $k$ contains \textit{small} classes}
    \State During the refinement of a class $C' \gets \{ j_1, j_2, j_3, \ldots j_{|C'|}\} \in C_k$ 
        \While{$j_m \in C' \mid m \in \{1,2,3\ldots|C'|\}$ is refined out of $C'$}
                \If{$prev_{C'}(j_m)$ exists}
            \State $j_{m-1} \gets prev_{C'}(j_m)$
            \If{$T[j_{m-1}] \ne k$ and $w[j_{m-1}..j_m + k - 1]$ is maximal right-closed}
                \State $insert(\mathcal{MRC}[j_{m-1}], (j_m + k - j_{m-1} ,k))$
            \EndIf
            \State $T[j_{m-1}] \gets k$
        \EndIf
        \If{$next_{C'}(j_m)$ exists}
            \State $j_{m+1} \gets next_{C'}(j_{m})$
            \If{$T[j_{m}] \ne k$ and $w[j_{m}..j_{m+1} + k - 1]$ is maximal right-closed}
                \State $insert(\mathcal{MRC}[j_{m}], (j_{m+1} + k - j_{m} ,k))$
            \EndIf
            \State $T[j_{m}] \gets k$
        \EndIf
    \EndWhile
    \State $k \gets k + 1$ 
    \State Compute \textit{small} classes at level $k$
\EndWhile
\State \Return $\mathcal{MRC}$
\EndProcedure
\end{algorithmic}
\end{algorithm}

Given a string $w[1..n]$, we first compute the maximal right-closed substrings of length $1$ and insert them into the $\mathcal{MRC}$ array, as shown in the \textsc{AddSingletonMRC} procedure presented in Algorithm~\ref{alg:addSingletons}. To compute all the maximal right-closed substrings of length greater than $1$ we adapt the CMR Algorithm to use an auxiliary \textit{boolean} array $T_k[1..n]$ at each level $k$, where $T_k[i]$ records whether $i$ has a successor in its class $C' \in C_k$. More precisely, for all $1 \le i \le n$, $T_k[i]=1 \Leftrightarrow$ ($\exists j \mid j = next_{C'}(i)$ and $C' \in C_k$); otherwise $T_k[i]=0$. Since a classes at level $k$ contain all the starting positions of a substrings of length $k$ grouped according to substring equality, $T_k[i]=1$ indicates the existence of a closed string at $i$.

At level $k$, $T_k$ is initialized to all zeros. Let $C' = \{ j_1, j_2, j_3, \ldots j_{|C'|}\}$ be an equivalence class such that $C' \in C_k$. Suppose, during refinement $C'$ is partitioned. Let $j_{m}$, where $m\in [1..|C'|]$, be the first index removed from $C'$. If $prev_{C'}(j_m)$ exists, let $j_{m-1} = prev_{C'}(j_m)$ and set $T_k[j_{m-1}] \gets k$, since $w[j_{m-1}..j_m+k-1]$ forms a closed substring. Similarly, if $next_{C'}(j_m)$ exists, let $j_{m+1} = next_{C'}(j_m)$ and set $T_k[j_m] \gets k$, since $w[j_m..j_{m+1}+k-1]$ forms a closed substring. In both cases, the border of the closed substring is $b=w[j_{m-1}..j_{m-1}+k-1]$ of length $k$. Recall that, by Fact~2 in Section~\ref{ssec:overviewCMR}, each class is represented as a doubly linked list with auxiliary pointers. Therefore, in Algorithm~\ref{alg:mrc_crochemore}, we use the $prev_{C'}()$ and $next_{C'}()$ pointers to simulate the behavior of an ordered set. 

To check if these closed substrings are maximal right-closed, we perform a constant time check by checking if $j_{m} + k = n$ or ($j_{m} + k < n$ and $w[j_{m-1}+k]\neq w[j_{m}+k]$), and $j_{m+1} + k = n$ or ($j_{m+1} + k < n$ and $w[j_{m}+k]\neq w[j_{m+1}+k]$), respectively. The lines $11$ and $16$ in Algorithm~\ref{alg:mrc_crochemore} perform these comparisons. If they are maximal right-closed, then we add them to the $\mathcal{MRC}$ array. 

Later in the refinement process, if $j_{m-1}$ is removed from $C'$, the longer string $w[j_{m-1}..j_{m+1}+k-1]$ with border $b$, also contains $b$ at $j_{m}$. Hence, it is not a closed substring, even though $j_{m+1} = next_{C'}(j_{m-1})$. Moreover, at any index position, we cannot have a closed string with two different borders of the same length. Therefore, the string $w[j_{m-1}..j_{m+1}+k-1]$ is either open or is closed with a longer border and would be found at a level greater than $k$ in the algorithm. 

The auxiliary array $T_k$ is used conceptually to determine whether a candidate substring is closed at level $k$. Maintaining a separate boolean array for each level would require $\mathcal{O}(n^2)$ space. To achieve $\Theta(n)$ space complexity, we instead maintain a single integer array $T[1..n]$, initialized to all zeros. The array $T$ acts as a level-dependent timestamp structure: the condition $T_k[i]=1$ is represented by $T[i]=k$. Thus, if a closed substring with border length $k$ is identified at position $i$ during level $k$, we set $T[i]\gets k$, as shown in lines $13$ and $18$ of Algorithm~\ref{alg:mrc_crochemore}, regardless of whether the corresponding substring is maximal right-closed. Consequently, whether a candidate substring starting at position $i$ has already been identified as closed at level $k$ can be determined in constant time by testing whether $T[i]=k$. A clear analysis of the $\Theta(n)$ space complexity of the CMR Algorithm is presented in Theorem 12.1.1~\cite{smyth2003computing}. Thus, we obtain the following result.

\begin{theorem}
\label{thm:mrc_crochemore_complexity}
Algorithm~\ref{alg:mrc_crochemore} correctly computes the $\mathcal{MRC}$ array of a string $w[1 .. n]$ without pre-processing in $\mathcal{O}(n \log n)$ time and $\Theta(n)$ auxiliary space.
\end{theorem}

\subsection{Algorithm to Compute \texorpdfstring{$\mathcal{C}(w)$}{C(w)} and MCSs}
Algorithm~\ref{alg:compact_representation} computes the compact representation $\mathcal{C}(w)$, as in Equation~\eqref{alg:compact_representation}, of all closed substrings of a string $w[1 .. n]$, by simply traversing all $n$ lists in the $\mathcal{MRC}$ array (see example in Table~\ref{tab:compact_repr_mississippi}). Since the $\mathcal{MRC}$ array is of size $\mathcal{O}(n \log n)$ we get Theorem~\ref{thm:C(w)_time_complexity_correctness}. 

\begin{algorithm}[htbp!]
\caption{Compute the compact representation $\mathcal{C}(w)$ using $\mathcal{MRC}$ array.}
\label{alg:compact_representation}
\begin{algorithmic}[1]
\Procedure{Compute-Compact-Representation}{$\mathcal{MRC}$}
\State $\mathcal{C} \gets \emptyset$
\For{$i \gets 1$ \textbf{to} $n$}
     \For{$j \gets 1$ \textbf{to} length of $\mathcal{MRC}[i]$}
        \State $(\ell r_j, \ell b_j) \gets \mathcal{MRC}[i][j]$
        \Comment{$\ell r_j=|r_j|, \ell b_j=|b_j|, \ell p_j=|p_j|$}
        \If{$j = 1$}
            \State $\ell p_j \gets 1$ \Comment{Equation~\eqref{eq:length_of_prefix}}
        \Else
            \State $(\ell r_{j-1}, \ell b_{j-1}) \gets \mathcal{MRC}[i][j-1]$
            \State $\ell p_j \gets \ell r_j - \ell b_j + \ell b_{j-1} + 1$ \Comment{Equation~\eqref{eq:length_of_prefix}}
        \EndIf
        \State insert$(\mathcal{C}, (i, \ell p_j, \ell r_j))$
    \EndFor
\EndFor
\State \Return $\mathcal{C}$ \Comment{$C$ is the compact representation as stated in Equation~\eqref{eq:compact_representation}}
\EndProcedure
\end{algorithmic}
\end{algorithm}

\begin{theorem}

\label{thm:C(w)_time_complexity_correctness}
Algorithm~\ref{alg:compact_representation} correctly computes all $\mathcal{O}(n^2)$ closed substrings of a string $w[1 .. n]$ in a compact representation in $\mathcal{O}(n \log n)$ time.
\end{theorem}

We show that the $\mathcal{MRC}$ array can also be used to compute all MCSs of a string $w[1 .. n]$ by performing a simple left to right scan of the array and conducting constant time checks of left extendibility, as presented in Algorithm~\ref{alg:mcs_using_mrc}. In particular, for each index $i$ in $w$ (where $1 \leq i \leq n$), the algorithm examines each maximal right-closed substring represented in $\mathcal{MRC}[i]$ and verifies whether the substring is also maximal left-closed using a constant time check $i = 1$ or $(i > 1 \text{ and } w[i - 1] \neq w[i + |r| - |b| - 1])$. If the condition is satisfied, the substring is added to the MCS list (see example in Table~\ref{tab:compact_repr_mississippi}). 

Scanning the $\mathcal{MRC}$ array requires at most $\mathcal{O}(n \log n)$ time, and each check for left extendibility is performed in constant time. Therefore, Algorithm~\ref{alg:mcs_using_mrc} computes all MCSs in $\mathcal{O}(n \log n)$ time, establishing Theorem~\ref{thm:mcs_complexity_correctness}.

\begin{algorithm}[htbp!]
\caption{Compute all MCSs of $w[1..n]$ using the pre-computed $\mathcal{MRC}$ array.}
\label{alg:mcs_using_mrc}
\begin{algorithmic}[1]
\Procedure{}{$w,\mathcal{MRC}$}
\State $mcsList \gets \emptyset$
\For{$i \gets 1$ \textbf{to} $n$}
    \For{$j \gets 1$ \textbf{to} length of $\mathcal{MRC}[i]$}
        \State $(\ell r, \ell b) \gets \mathcal{MRC}[i][j]$ \Comment{$\ell r=|r|, \ell b=|b|$}
        \If{$i = 1$ \textbf{or} $(i > 1 \textbf{ and } w[i - 1] \neq w[i + \ell r - \ell b - 1])$}
            \State $insert(mcsList, w[i .. i + \ell r - 1])$
        \EndIf
    \EndFor
\EndFor
\State \Return $mcsList$ \Comment{$mcsList$ contains all MCSs of $w[1..n]$}
\EndProcedure
\end{algorithmic}
\end{algorithm}

\begin{theorem}
\label{thm:mcs_complexity_correctness}
Algorithm~\ref{alg:mcs_using_mrc} correctly computes all MCSs of a string $w[1 .. n]$ in $\mathcal{O}(n \log n)$ time.
\end{theorem}

Note that Badkobeh et al.~\cite{BADKOBEH2026115628} also achieve the same theoretical time complexity for computing MCSs. Their approach relies on a bottom-up traversal of a binary suffix tree~\cite{Weiner73} (i.e., a suffix tree transformed into one with at most two children) for $w[1..n]$ to compute all MCSs. In contrast our method employed in the $\mathcal{MRC}$ array construction (Algorithm~\ref{alg:mrc}), exploits the rise and fall of the LCP array values, which are tracked using a stack that effectively simulates the behavior of a bottom-up tree traversal to compute all maximal right closed substrings.
Furthermore, although both approaches employ AVL trees to store and merge sorted lists, they differ in how these trees are augmented. This distinction arises because our work and that of Badkobeh et al.\ build on different results of Brodal et al.—namely, quasiperiodicity~\cite{brodal2000} and gapped repeats~\cite{BLPS99}, respectively. Consequently, the augmentation schemes differ, leading to structurally distinct AVL trees. In addition, we employ an adapted version of the augmented AVL tree of Brodal et al., as refined by Kociumaka et al.~\cite{Jakub2015}, incorporating the \textsc{Union} and \textsc{ChangeList} operations. As a result, the underlying data structures differ significantly, which in turn leads to non-trivial differences in the complexity analysis.

Finally, the main contribution of Algorithm~\ref{alg:mrc} is the efficient computation of the $\mathcal{MRC}$ array, a data structure that facilitates the identification of all maximal right-closed strings, rather than all MCSs, in order to obtain a compact representation of all closed substrings—a result not presented in~\cite{BADKOBEH2026115628}. Algorithm~\ref{alg:mcs_using_mrc} is then a straightforward application of this data structure.

We were unable to experimentally compare the performance of Algorithm~\ref{alg:mcs_using_mrc} with that of~\cite{BADKOBEH2026115628}. Nevertheless, we expect our implementation to perform better in practice, particularly for long strings, due to its conceptual simplicity and the fact that it avoids the use of the space-intensive suffix tree.

\begin{table}[h]
\centering
\renewcommand{\arraystretch}{1.2}
\begin{tabular}{>{\centering\arraybackslash}m{2cm} >{\raggedright\arraybackslash}m{4cm} >{\raggedright\arraybackslash}m{3cm} >{\centering\arraybackslash}m{2cm}}
\toprule
\textbf{$C(w)$} & \textbf{All Closed Factors} & \textbf{Maximal Right Closed Factors} & \textbf{MCS?} \\
\midrule
$(1,1,1)$ & $m$ & $m$ & Yes \\
$(2,4,7)$ & $issi$, $issis$, $ississ$, $ississi$ & $ississi$ & Yes \\
$(2,1,1)$ & $i$ & $i$ & Yes \\
$(3,5,6)$ & $ssiss$, $ssissi$ & $ssissi$ & No \\
$(3,1,2)$ & $s$, $ss$ & $ss$ & Yes \\
$(4,5,5)$ & $sissi$ & $sissi$ & No \\
$(4,3,3)$ & $sis$ & $sis$ & Yes \\
$(4,1,1)$ & $s$ & $s$ & No \\
$(5,4,4)$ & $issi$ & $issi$ & No \\
$(5,1,1)$ & $i$ & $i$ & Yes \\
$(6,1,2)$ & $s$, $ss$ & $ss$ & Yes \\
$(7,1,1)$ & $s$ & $s$ & No \\
$(8,4,4)$ & $ippi$ & $ippi$ & Yes \\
$(8,1,1)$ & $i$ & $i$ & Yes \\
$(9,1,2)$ & $p$, $pp$ & $pp$ & Yes \\
$(10,1,1)$ & $p$ & $p$ & No \\
$(11,1,1)$ & $i$ & $i$ & Yes \\
\bottomrule
\end{tabular}
\caption{Compact representation $\mathcal{C}(w)$ of $w[1..11] = mississippi$ with corresponding closed factors, maximal right-closed factors, and MCSs.}
\label{tab:compact_repr_mississippi}
\end{table}

\section{MCSs in a Fibonacci Word}
\label{sec:enomifs}

In this section, we study the occurrences of MCSs in a Fibonacci word ($f_n$), and provide a formula to compute the exact number of MCSs in $f_n$. Note that throughout this section, $n$ refers to the $n$-th Fibonacci number or word, rather than to the string length.

Let us recall that a \df{Fibonacci} word $f_n$ is defined recursively by $f_0 = 0$, $f_1 = 1$, and $f_n = f_{n-1}f_{n-2}$ for $n \geq 2$. We denote $F_n = |f_n|$, where $F_n$ satisfies the Fibonacci recurrence relation $F_n = F_{n-1} + F_{n-2}$, with $F_0 = 1$ and $F_1 = 1$. We refer to the logical separation between $f_{n-1}$ and $f_{n-2}$, in the string $f_n$, as the \df{boundary}.

Bucci et al.~\cite{BUCCI201312} study the open and closed prefixes of Fibonacci words and show that the sequence of open and closed prefixes of a Fibonacci word follows the Fibonacci sequence. De Luca et al.~\cite{DELUCA201727} explore the sequence of open and closed prefixes of a Sturmian word. In~\cite{JAHANNIA202232}, the authors define and find the closed Ziv--Lempel factorization and classify closed prefixes of infinite $m$-bonacci words. In this section, we extend previous work on closed substrings in Fibonacci words by deriving a formula to compute the number of MCSs they contain.

For any integer $n \geq 2$, let $p_n = f_n[1..F_n - 2]$ denote the prefix of $f_n$ excluding its last two symbols. Then, $f_n = p_n \, \delta_n$, where $\delta_n = 10$ if $n$ is even and $\delta_n = 01$, otherwise. Let $P_n = F_n - 2$ denote the length of $p_n$. The golden ratio is defined as $\phi = \frac{1 + \sqrt{5}}{2} $, and it is known that the ratio of consecutive Fibonacci numbers converges to $\phi$, i.e., $\lim_{n \to \infty} \frac{F_{n}}{F_{n-1}} = \phi.$ 

The non-singleton MCSs are either runs or maximal gapped repeats in which the arm occurs exactly twice~\cite{Badkobeh_2022_SPIRE}. For a Fibonacci word $f_n$, let $SM(f_n)$, $R(f_n)$, and $GM(f_n)$ denote the number of singleton MCSs, runs, and gapped-MCSs, respectively. Therefore, the total number of MCSs in $f_n$ is given by $M(f_n) = SM(f_n) + R(f_n) + GM(f_n)$.

\begin{lemma}
\label{lem:sm_fib}
The no. of singleton MCSs in a Fibonacci word $f_n$, for $n \geq 4$, is:
\begin{equation}
SM(f_n) = 
\begin{cases} 
F_{n-2} + F_{n-4} + 2, & \text{if } n \text{ is odd,} \\
F_{n-2} + F_{n-4}, & \text{if } n \text{ is even.}
\end{cases}
\label{eq:sm}
\end{equation}
\end{lemma}

\begin{proof}
We prove the Lemma by induction on $n$. The base cases, $SM(f_4) = 3$ and $SM(f_5) = 6$, match the direct counts. 

For the inductive step, assume that Equation~\eqref{eq:sm} holds for $k \geq 6$. By definition, $f_{k+1} = f_k f_{k-1}$. Every Fibonacci word $f_k$ begins with a $10$; if $k$ is odd, it ends with $01$ and if $k$ is even, it ends with $10$. By definition, the first and last character in $f_k$ are singleton MCSs. Therefore, we examine whether the singleton MCSs at the boundary are retained in $f_{k+1}$. 

Suppose $k+1$ is even. By inductive hypothesis, $SM(f_k) = F_{k-2} + F_{k-4} + 2$ and $SM(f_{k-1}) = F_{k-3} + F_{k-5}$. Since $k$ is odd, $f_k$ ends with $1$ and $f_{k-1}$ begins with $1$, their concatenation results in the string $11$ at the boundary, which forms a new non-singleton MCS, and reduces the count of the singleton MCSs in $f_{k+1}$ by two. Thus, we get $SM(f_{k+1}) = SM(f_k) + SM(f_{k-1}) - 2 = F_{k-1} + F_{k-3}$.

Suppose $k+1$ is odd. By inductive hypothesis, $SM(f_k) = F_{k-2} + F_{k-4}$ and $SM(f_{k-1}) = F_{k-3} + F_{k-5} + 2$. Since $k$ is even, $f_k$ ends with $0$ and $f_{k-1}$ begins with $1$, their concatenation results in the string $01$ at the boundary, which does not reduce the count of singleton MCSs in $f_{k+1}$. Thus, we get $SM(f_{k+1}) = SM(f_k) + SM(f_{k-1}) = F_{k-1} + F_{k-3} + 2$. 
\end{proof}

\begin{lemma}[Theorem~2 in~\cite{kk_runs_in_fib}]
\label{lem:runs_in_fn}
For $n \geq 4$, the number of runs in the Fibonacci word $f_n$, is given by $R(f_n) = 2F_{n-2} - 3$.
\end{lemma}

To compute $GM(f_n)$, we count the number of maximal gapped repeats whose arms occur exactly twice, as these are, by definition gapped-MCSs. In Theorem~\ref{thm:gapped_mcs_fn_unique} we first show that the only gapped-MCSs in a Fibonacci word are occurrences of the substring $101$ that are also maximal gapped repeats. Then, we count all such occurrences in $f_n$.

\begin{lemma}[Theorem~1 in~\cite{SOFSEM_25_maximal_gapped_repeats_fn}]
\label{lem:suffix_gapped_repeat}
Suppose $u v u$ is a maximal gapped repeat in $f_n$ that is also a suffix of $f_n$. Then, the arm $u$ is a Fibonacci word.
\end{lemma}

\begin{lemma}
\label{lem:suffix_gapped_mcs}
For $n \geq 5$, suppose $u v u$ is a maximal gapped repeat in $f_n$ that is also a suffix of $f_n$. Then $u v u$ is not a gapped-MCS.
\end{lemma}
\begin{proof}
By Lemma~\ref{lem:suffix_gapped_repeat}, the arm of the maximal gapped repeat is a Fibonacci word $u = f_k$. Clearly, $k \ne n$. 

If $n$ is even, then the proper suffixes of $f_n$ that are Fibonacci words are in the set $\mathcal{S}=\{f_{k} \mid k \mod 2 = 0 \land 0\leq k < n\}$. For any proper suffix $f_k \in \mathcal{S} \setminus \{f_0, f_2\}$, $f_{k+2}$ is a suffix of $f_n$. By definition, $f_{k+2} = f_{k+1} f_{k}=f_k f_{k-1} f_k = f_k f_k f_{k-3} f_{k-2}$, the suffix $f_{k-1} f_k = f_k f_{k-3} f_{k-2}$ has length $F_k + F_{k-1} < 2F_k$ implying that the last two occurrences of $f_k$ overlap. Therefore, any $u v u$ with $f_k$ as its arm, will have at least three occurrences of $f_k$ making it an open string. For suffix $f_k \in \{f_0, f_2\}$ of $f_n$, for $n \geq 6$, $f_6 = 1011010110110$ is a suffix of $f_n$, and the last two occurrences of $f_k$ are left-extendible, and so, it is not maximal. Moreover, any longer $uvu$ with the arm $f_k$ will have at least three occurrences of $f_k$ making it an open string. 

If $n$ is odd, then $\mathcal{S}=\{f_{k} \mid k \mod 2 = 1 \land 
1\leq k < n\}$. The argument is analogous for $f_k \in \mathcal{S} \setminus \{f_1\}$ and $f_k \in \{f_1\}$, respectively. 
\end{proof}

\begin{lemma}[Theorem~2 in~\cite{SOFSEM_25_maximal_gapped_repeats_fn}]
\label{lem:non_suffix_gapped_repeat}
For $n \geq 3$, suppose $uvu$ is a maximal gapped repeat in $f_n$ that is not a suffix of $f_n$. Then the arm $u$ belongs to the set $\mathcal{A}_n = \{p_3, p_4, \ldots, p_{n-2}\}$.
\end{lemma}

\begin{lemma}[Lemma~8 in~\cite{SOFSEM_25_maximal_gapped_repeats_fn}]
\label{lem:borders_of_p_n}
For $n \geq 4$, all the borders of $p_n$ form the set $\mathcal{B}_n = \{p_3, p_4, \ldots, p_{n-1}\}$.
\end{lemma}

\begin{lemma}
\label{lem:longest_cover_of_p_n}
For $n \geq 5$, the border $p_{n-1}$ of $p_n$ is the longest proper cover of $p_n$.
\end{lemma}

\begin{proof}
By Lemma~\ref{lem:borders_of_p_n}, we know that $p_{n-1}$ is the longest border of $p_n$. $F_n=F_{n-1} + F_{n-2}$. For $n\geq 5$, $F_n-4=F_{n-1} + F_{n-2}-4$, and so $P_n-2=P_{n-1} + P_{n-2}$. Hence, $P_n \leq 2 \cdot P_{n-1}$. Clearly, $p_{n-1}$ is either an adjacent or an overlapping border of $p_n$. Therefore, $p_{n-1}$ is the longest proper cover of $p_n$.
\end{proof}

\begin{lemma}[Lemma 2 in~\cite{MOORE1994239}]
\label{lem:cover_inheritance}
Let $u$ be a proper cover of $x$ and let $v \ne u$ be a substring of $x$ such that $|v| \leq |u|$. Then, $v$ is a cover of $x \Leftrightarrow v$ is a cover of $u$.
\end{lemma}

\begin{lemma}
\label{lem:covers_of_p_n}    
For $n \geq 5$, all covers of $p_n$ form the set $\mathcal{C}_n = \{p_4, p_5, \ldots, p_n\}$.
\end{lemma}

\begin{proof}
By Lemma~\ref{lem:longest_cover_of_p_n}, for $n \geq 5$, $p_{n-1}$ is the longest proper cover of $p_n$. Similarly, $p_{n-2}$ is the longest proper cover of $p_{n-1}$, $p_{n-3}$ is the longest proper cover of $p_{n-2}$, and so on. Therefore, by Lemma~\ref{lem:cover_inheritance}, it follows that all covers of $p_n$ belong to the set $\mathcal{C}_n = \{p_4, p_5, \ldots, p_n\}$ for all $n \geq 5$. 
\end{proof}

\begin{theorem}
\label{thm:gapped_mcs_fn_unique}
For $n \geq 5$, every gapped-MCS in a Fibonacci word $f_n$ corresponds to an occurrence of the substring $101$ that is also a maximal gapped repeat.
\end{theorem}
\begin{proof}
All gapped-MCSs are maximal gapped repeats. By Lemma~\ref{lem:suffix_gapped_mcs}, for $n \geq 5$, all maximal gapped repeats that are also suffixes of $f_n$ are not gapped-MCSs.

By Lemma~\ref{lem:non_suffix_gapped_repeat}, for $n \geq 3$, if $u v u$ is a maximal gapped repeat in $f_n$ that is not a suffix of $f_n$, then the arm $u$ belongs to the set $\mathcal{A}_n = \{p_3, p_4, \ldots, p_{n-2}\}$. By Lemma~\ref{lem:covers_of_p_n}, all covers of $p_n$ belong to the set $\mathcal{C}_n = \{p_4, p_5, \ldots, p_n\}$. 

Suppose there exists a maximal gapped $w=uvu$, where $u \in \mathcal{A}_n \setminus {p_3}$, that is a gapped-MCS. Assume $n$ is odd. Then, $f_n = p_n \delta_n$, where $\delta_n=01$, and $f_n[F_n - 1] = 0$. Since $u \in \mathcal{A}_n \setminus {p_3}$, it ends with $1$. Clearly, $u$ does not occur ending at position $F_n - 1$. Now suppose $n$ is even. By a similar reasoning, $u$ does not occur ending at $F_n - 1$ since it ends with $01$ and $f_n[F_n - 2] = f_n[F_n - 1] = 1$ (as $f_4 = 10110$ is a suffix of $f_n$). Hence, $u$ ends within $p_n$. Since $u$ is a cover of $p_n$, $w=uvu$ will contain at least three occurrences of $u$ making it an open string. Therefore, there is no maximal gapped repeat of the form $w=uvu$, where $u \in \mathcal{A}_n \setminus {p_3}$, that is a gapped MCS.

Now we consider the final case, where $u = p_3 = 1$. In this case, we get $101$ as the only gapped-MCS. Any longer maximal gapped repeat will have at least three occurrences of $u=p_3 = 1$, making it an open string --- a contradiction. 
\end{proof}

\begin{lemma}
\label{lem:gapped_mcs_fn}
The no. of gapped-MCSs in a Fibonacci word $f_n$, for $n \geq 5$, is:
\begin{equation}
GM(f_n) =
\begin{cases}
F_{n-5}, & \text{if } n \text{ is odd}, \\
F_{n-5} + 1, & \text{if } n \text{ is even}.
\end{cases}
\label{eq:gapped_mcs_fn}
\end{equation}
\end{lemma}
\begin{proof}
We prove the Lemma by induction on $n$. By direct counting for the base cases, we verify that $GM(f_5) = 1$ and $GM(f_6) = 2$. For the inductive step, assume that Equation~\eqref{eq:gapped_mcs_fn} holds for $k \geq 7$. For $n \geq 3$, $f_3 = 101$ is a prefix of every $f_n$, and by Theorem~\ref{thm:gapped_mcs_fn_unique} every gapped-MCS in a Fibonacci word $f_n$, for $n \geq 5$, corresponds to an occurrence of the substring $101$ that is also a maximal gapped repeat. In $f_{k+1} = f_k f_{k-1}$, any new occurrences of $101$ in $f_{k+1}$ can only occur beginning in $f_k$ and ending in $f_{k-1}$. Moreover, an occurrence of $101$ as a suffix of $f_k$ or a prefix of $f_{k-1}$ may be lost by extension. Below we examine all such cases. 

Suppose $k+1$ is odd. By inductive hypothesis, $GM(f_k) = F_{k-5} + 1$ and $GM(f_{k-1}) = F_{k-6}$. Moreover, $f_k$ ends with the suffix $f_2= 10$ and $f_{k-1}$ begins with $f_3=101$, resulting in the formation of the string $101$ crossing the boundary. Clearly, this string is not a gapped-MCS as it is right-extendible. The gapped-MCS $101$ which is the prefix of $f_{k-1}$ is no longer a gapped-MCS since $f_k$ ends with a $0$ making it left-extendible. Thus, we get $ GM(f_{k+1}) = GM(f_k) + GM(f_{k-1}) - 1 = (F_{k-5} + 1) + F_{k-6} - 1 = F_{k-4}$. 

\sloppy Suppose $k+1$ is even. By inductive hypothesis, $GM(f_k) = F_{k-5}$ and $GM(f_{k-1}) = F_{k-6}+1$. $f_k$ ends with a $1$, so the prefix $101$ of $f_{k-1}$ is retained as a gapped-MCS. By Lemma~\ref{lem:suffix_gapped_mcs} the suffix $f_3 = 101$ of $f_k$ is not a gapped-MCS. Thus, we get $GM(f_{k+1}) = GM(f_k) + GM(f_{k-1}) = F_{k-4} + 1$. 
\end{proof}

\begin{theorem}
The number of MCSs in a Fibonacci word $f_n$, denoted by $M(f_n)$, for $n \geq 5$, is given below in Equation~\eqref{eq:MCS_count}. Furthermore, $M(f_n) \approx 1.382 F_n$.
\begin{equation}
M(f_n) = 
\begin{cases} 
F_n + F_{n-2} - 1, & \text{if } n \text{ is odd,} \\
F_n + F_{n-2} - 2, & \text{if } n \text{ is even.}
\end{cases}
\label{eq:MCS_count}
\end{equation}
\end{theorem}

\begin{proof}
Adding Equations~\eqref{eq:sm}, ~\eqref{eq:gapped_mcs_fn} and the Equation in Lemma~\ref{lem:runs_in_fn} and simplifying, we get Equation~\eqref{eq:MCS_count}. Next, we have $M(f_n) < F_n + F_{n-2} = \left(1 + \frac{F_{n-2}}{F_n}\right) F_n \approx \left(1 + \frac{1}{\phi^2}\right) F_n \approx 1.382 F_n$ (since $\lim_{n \to \infty} \frac{F_{n-2}}{F_n} = \frac{1}{\phi^2}$). 
\end{proof}

\section{Experimental Evaluation}
\label{sec:experimental_eval}

In this section, we evaluate the performance of the two proposed algorithms— Algorithm~\ref{alg:mrc} ($\mathsf{SA}$ \& $\mathsf{LCP}$ based) and Algorithm~\ref{alg:mrc_crochemore} (CMR based)—on different types of strings: periodic, aperiodic, DNA and random. We also exhaustively generate all strings over small alphabets of size $\sigma = 2, 3,$ and $4$ and of length $n$ for $1 \leq n \leq 20$, and compute the maximum number of maximal closed substrings (MCSs) they contain.

\subsection{Implementation Details}
\label{sec:implementation_details}
The code was implemented in C++ and compiled using GCC 11.5.0 with \texttt{-O3} flag and is available at \url{https://github.com/neerjamhaskar/closedStrings}.

The implementations for computing the suffix array ($\mathsf{SA}$) and the longest common prefix array ($\mathsf{LCP}$) were sourced from the \textit{libsais} library~\cite{Grebnov:libsais}, which builds on the contributions of~\cite{lcp2009,nong2011two,sais2014,suffixSort2020}.

The AVL trees used in Algorithm~\ref{alg:mrc} are implemented using the C++ \texttt{std::set} container. To maintain the logarithmic balancing properties of the AVL trees, the \texttt{std::set} container is implemented as a red-black tree, providing equivalent asymptotic guarantees and efficient memory management as the AVL trees. The \texttt{std::set} interface supports key operations such as \texttt{insert}, \texttt{set::upper\_bound}, and \texttt{set::lower\_bound}, all of which operate in $\mathcal{O}(\log n)$ time. These operations are particularly useful for efficiently computing the $\textsc{ChangeList}$, which records updates to successor and predecessor relationships between sets during \textsc{Union} operations. Specifically, the \texttt{set::insert} operation merges elements into a set, while \texttt{set::upper\_bound} and \texttt{set::lower\_bound} enable fast determination and update of adjacency relationships within the merged sets. Note that, our independently developed \textsc{ChangeList()} implementation, is closely related to the implementation of partial covers available at \url{https://github.com/heurezjusz/Quasiperiods}, developed for~\cite{CZAJKA202117}.

A simple stack is employed to store pointers to the suffix sets being processed at various stages of Algorithm~\ref{alg:mrc}, allowing efficient access to both current and previously processed sets with minimal memory overhead. The implementation uses the C++ \texttt{std::stack} template to ensure type safety and flexibility during execution.

For Algorithm~\ref{alg:mrc_crochemore}, the implementation of the CMR Algorithm for computing maximal repetitions was sourced from~\cite{Franek2002Crochemore}. We modified this implementation to compute the $\mathcal{MRC}$ array. In both of our algorithms, the $\mathcal{MRC}$ array is implemented as \texttt{std::vector} of \texttt{std::list}, enabling constant-time access and insertion to either ends of the list.

\subsection{Experiments}
\label{sec:experiments}
All experiments were carried out on a server with an Intel Xeon Gold 6426Y CPU (64 cores, 128 threads, 75 MiB L3 cache), 250 GiB RAM, running RHEL 9.5 (kernel version \texttt{5.14.0-503.26.1}). In our experiments the running time of Algorithm~\ref{alg:mrc} includes the pre-processing required to construct the suffix array ($\mathsf{SA}$) and longest common prefix array ($\mathsf{LCP}$) using the \texttt{libsais} library.

\begin{figure}[htbp!]
    \centering
    \includegraphics[width=1.0\textwidth]{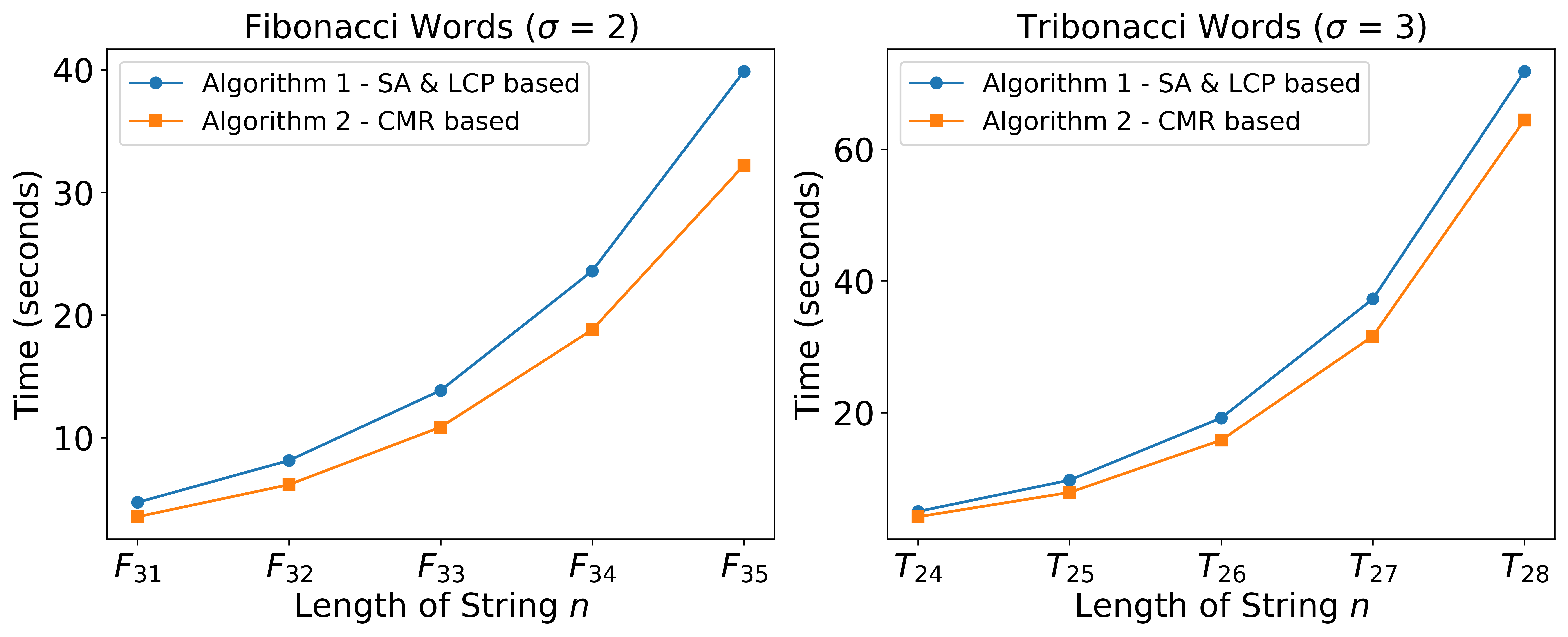}
    \caption{Execution time comparison between Algorithm~\ref{alg:mrc} ($\mathsf{SA}$ \& $\mathsf{LCP}$ based) and Algorithm~\ref{alg:mrc_crochemore} (CMR based) on Fibonacci and Tribonacci words.}
    \label{fig:periodic}
\end{figure}

We first run Algorithm~\ref{alg:mrc} and Algorithm~\ref{alg:mrc_crochemore} on well-studied Fibonacci and Tribonacci words which are known to be highly periodic. The Tribonacci word $t_n$ is defined recursively as $t_n = t_{n-1} t_{n-2} t_{n-3}$ with $t_0 = 1$, $t_1 = 12$, and $t_2 = 1213$; the length of each Tribonacci word is denoted as $T_n = |t_n|$. As shown in Figure~\ref{fig:periodic}, Algorithm~\ref{alg:mrc_crochemore} consistently outperforms Algorithm~\ref{alg:mrc}.

\begin{figure}[htbp!]
    \centering
    \includegraphics[width=1.0\textwidth]{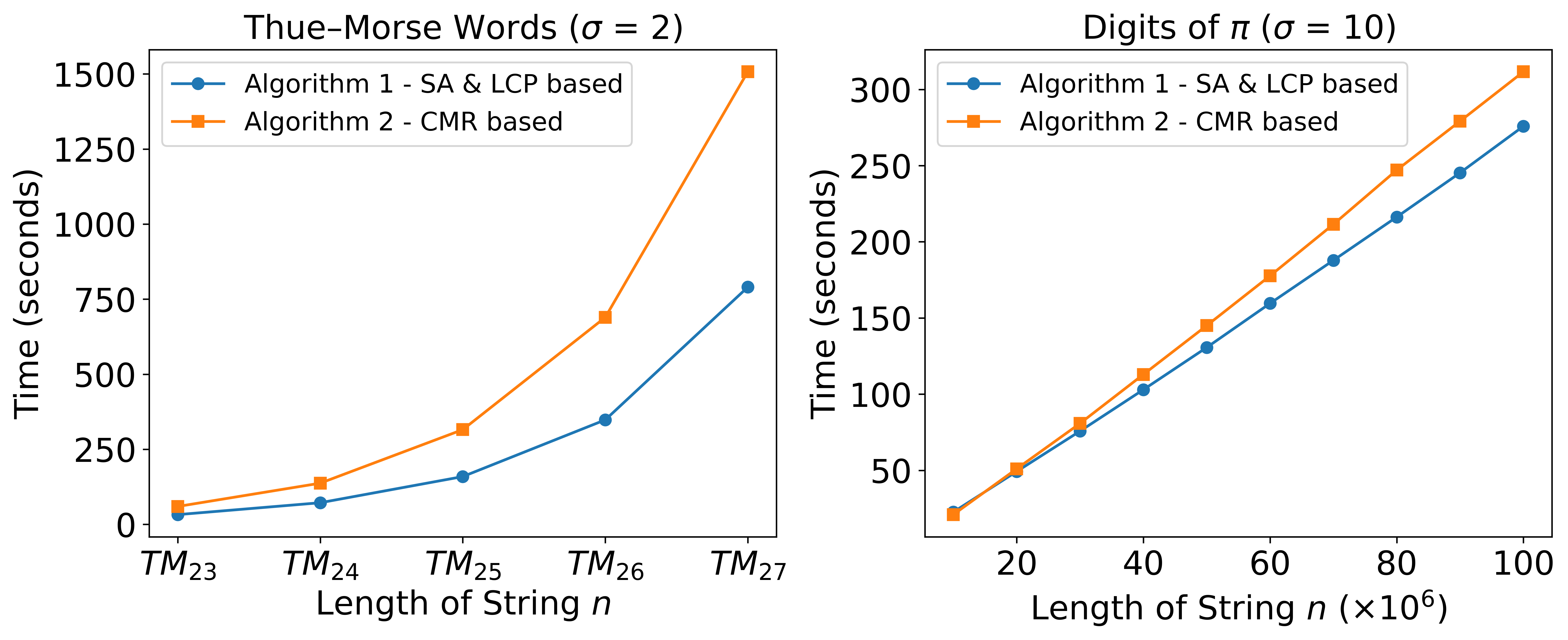}
    \caption{Execution time comparison between Algorithm~\ref{alg:mrc} ($\mathsf{SA}$ \& $\mathsf{LCP}$ based) and Algorithm~\ref{alg:mrc_crochemore} (CMR based) on aperiodic Thue-Morse words and digits of $\pi$.}
    \label{fig:tm_pi}
\end{figure}

We then evaluate Algorithm~\ref{alg:mrc} and Algorithm~\ref{alg:mrc_crochemore} on Thue--Morse words and the digits of $\pi$ (which have lesser number of periodic substrings and are non-repetitive). The Thue--Morse words $tm_n$ are generated for our experiments using a Python implementation of the standard morphism $\mu(0)=01$ and $\mu(1)=10$ with $tm_0 = 0$; the length of each word is $TM_n = |tm_n| = 2^{n}$. The digits of $\pi$ are regarded as an effectively aperiodic sequence for computational purposes. As shown in Figure~\ref{fig:tm_pi}, Algorithm~\ref{alg:mrc} outperforms Algorithm~\ref{alg:mrc_crochemore} on these string classes.

\begin{figure}[htbp!]
    \centering
    \includegraphics[width=1.0\textwidth]{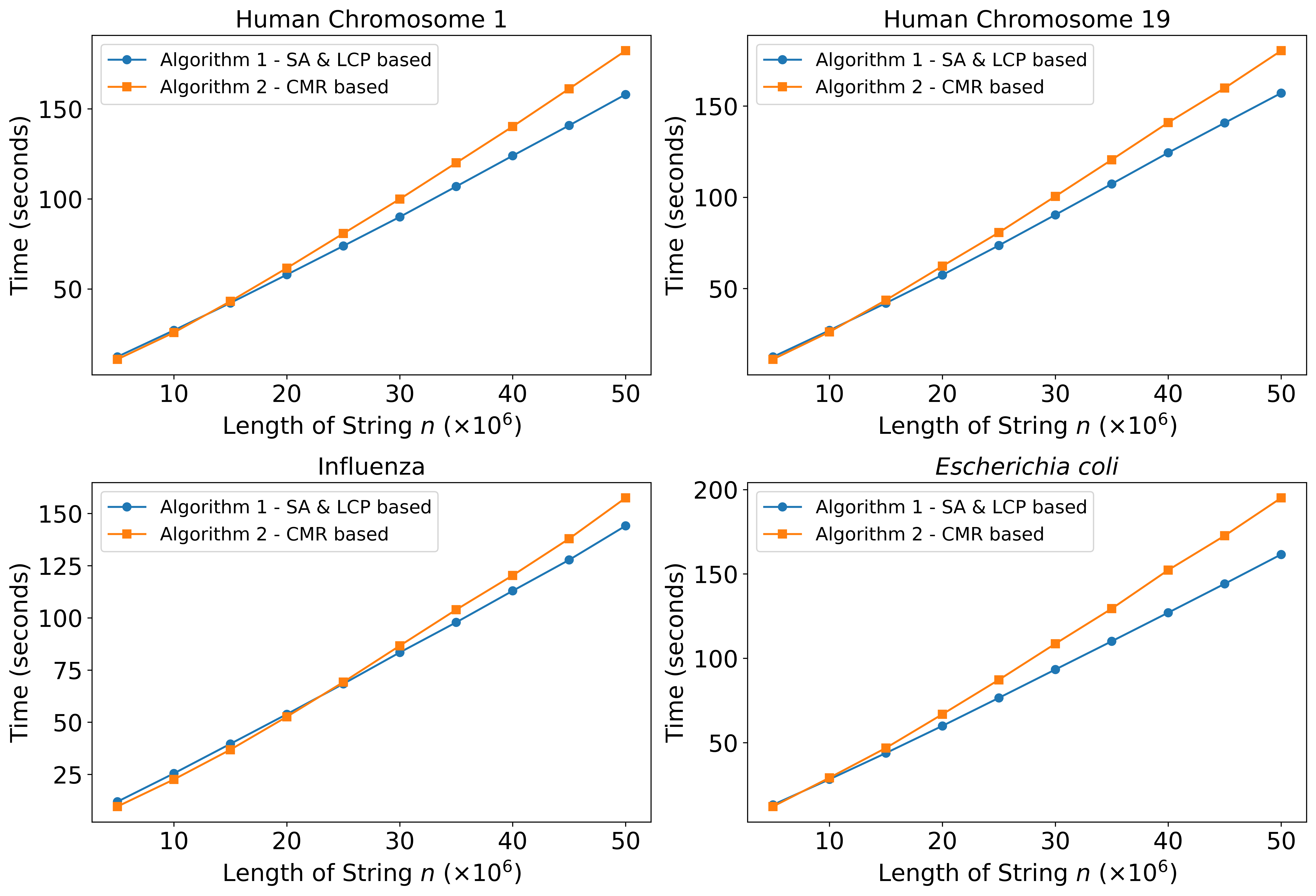}
    \caption{Execution time comparison between Algorithm~\ref{alg:mrc} ($\mathsf{SA}$ \& $\mathsf{LCP}$ based) and Algorithm~\ref{alg:mrc_crochemore} (CMR based) on DNA sequences--prefixes of Human Chromosome~$1$, Human Chromosome~$19$, \textit{Escherichia coli}, and Influenza.}
    \label{fig:dna}
\end{figure}

For our experiments on biological sequences, we evaluate both Algorithm~\ref{alg:mrc} and Algorithm~\ref{alg:mrc_crochemore} on four DNA datasets: prefixes of Human Chromosome~$1$, Human Chromosome~$19$, \textit{Escherichia coli}, and Influenza. Each dataset consists of increasingly larger prefixes, from $5$M up to $50$M, in increments of $5$M characters. Across all four datasets, Algorithm~\ref{alg:mrc} consistently outperforms Algorithm~\ref{alg:mrc_crochemore}, with the performance gap widening as the string length increases as shown in Figure~\ref{fig:dna}.

Next, we generate random strings over alphabets of size $2$, $4$, and $20$ in increments of $10$M, from $10$M up to $100$M, using the \textit{Pizza\&Chili} (\url{https://pizzachili.dcc.uchile.cl/experiments.html}) random string generator. For each string, we run both Algorithm~\ref{alg:mrc} and Algorithm~\ref{alg:mrc_crochemore} three times and report the median running time. Across all string lengths and alphabet sizes, we observe that Algorithm~\ref{alg:mrc} consistently outperforms Algorithm~\ref{alg:mrc_crochemore}.

\begin{figure}[htbp!]
    \centering
    \includegraphics[width=1.0\textwidth]{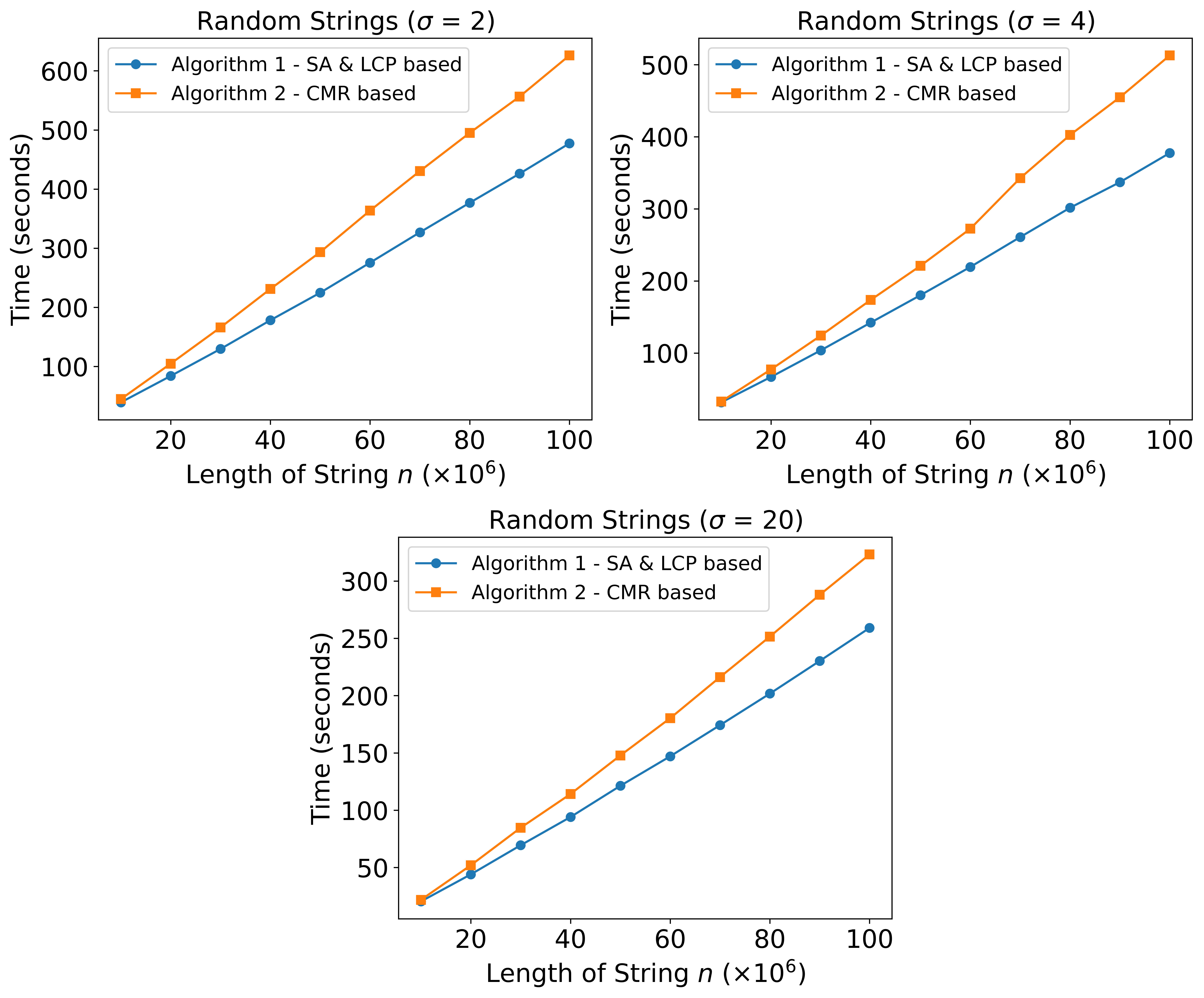}
    \caption{
        Execution time comparison between Algorithm~1 ($\mathsf{SA}$ \& $\mathsf{LCP}$ based) and Algorithm~2 (CMR based) on random strings.
    }
    \label{fig:spire_vs_crochemore_sigma_comparison}
\end{figure}

Overall, although Algorithm~\ref{alg:mrc_crochemore} is slower than Algorithm~\ref{alg:mrc} on most string classes considered, it offers greater potential for parallelization. In Algorithm~\ref{alg:mrc}, parallelism is largely limited to the initial partition of suffixes into at most $\sigma$ sets. In contrast, as Algorithm~\ref{alg:mrc_crochemore} progresses through successive levels, the refinement process typically produces an increasing number of equivalence classes, thereby enabling a higher degree of parallel execution.

\begin{figure}[ht!]
    \centering
    \includegraphics[width=0.75\linewidth]{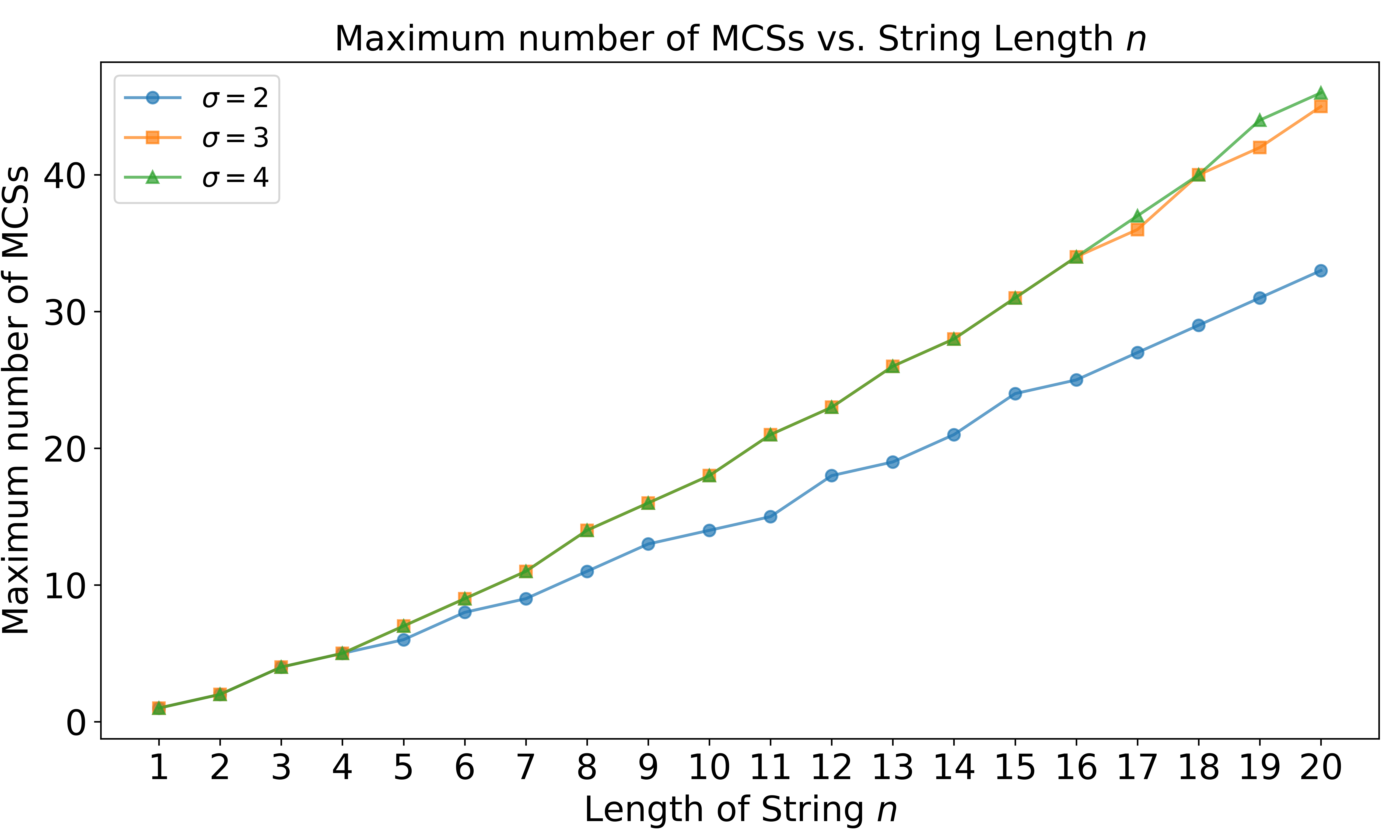}
    \caption{Maximum number of MCSs in strings of length $n$, where $1 \leq n \leq 20$, for alphabets of size $\sigma = 2, 3$ and $4$.}
    \label{fig:max_mcs_vs_n}
\end{figure}

For $\sigma = 2,3$ and $4$, we generate all strings of length $n$ for $1 \leq n \leq 20$ and compute their MCSs. The results are plotted in Figure~\ref{fig:max_mcs_vs_n}. As observed, the maximum number of MCSs increases with the size of the alphabet $\sigma$ for strings of the same length. In total, the number of strings analyzed is equal to $\sum_{i=1}^{20} (2^i + 3^i + 4^i) \approx 1.47$ trillion. While the relatively small range of lengths considered makes it difficult to determine the precise mechanism by which $\sigma$ affects the number of MCSs, the increasing separation between the curves in Figure~\ref{fig:max_mcs_vs_n} suggests that the alphabet size is indeed a contributing factor to the maximum number of MCSs in a string of length $n$, in part because the number of singleton MCSs increase with $\sigma$.

\section{Conclusion and Open Problems}
\label{sec:conclusion_and_open_problems}

We present a compact representation of all $\mathcal{O}(n^2)$ closed substrings of a string $w[1..n]$, requiring only $\mathcal{O}(n \log n)$ space. To compute this representation and identify all maximal closed substrings (MCSs), we introduce the $\mathcal{MRC}$ array and propose two $\mathcal{O}(n \log n)$ time algorithms (Algorithm~\ref{alg:mrc} and Algorithm~\ref{alg:mrc_crochemore}) for its computation. Algorithm~\ref{alg:mrc} requires pre-processing of the input string to compute the $\mathsf{SA}$ and $\mathsf{LCP}$ arrays, whereas Algorithm~\ref{alg:mrc_crochemore} does not require pre-processing and is based on the equivalence-class approach of the CMR algorithm. We also present an exact formula for computing the number of MCSs in Fibonacci words. Finally, we provide implementations and experimental analysis of the performance of Algorithm~\ref{alg:mrc} and Algorithm~\ref{alg:mrc_crochemore}. Our results show that Algorithm~\ref{alg:mrc} performs best for most string classes, while Algorithm~\ref{alg:mrc_crochemore} outperforms Algorithm~\ref{alg:mrc} for highly periodic strings.

Several questions remain open for future research. For example, Algorithm~\ref{alg:mrc} allows parallelization only up to $\sigma$, whereas Algorithm~\ref{alg:mrc_crochemore} can be parallelized more aggressively; determining whether this advantage leads to better practical performance is an interesting challenge. Another intriguing direction is the algorithmic generation of strings that attain the maximum number of MCSs for arbitrary $n$ and alphabet size $\sigma$. Moreover, extending the characterization of MCSs to $m$-bonacci words or more general Sturmian sequences presents a significant theoretical challenge. A deeper understanding of their structural and enumerative properties could lead to broader insights into the combinatorics of words and the design of efficient string algorithms.

\backmatter

\bmhead{Acknowledgements}
We thank Simon J. Puglisi for introducing us to the MCS problem using $\mathsf{SA}$ and $\mathsf{LCP}$, which led to Theorem~\ref{thm:mcs_complexity_correctness}. We are grateful to the reviewers for their valuable feedback.

\section*{Declarations}

\bmhead{Funding}
Neerja Mhaskar was funded by Natural Sciences \& Engineering Research Council of Canada [Grant Number RGPIN-2024-06915].

\bmhead{Competing interests}
The authors declare no competing interests.

\bmhead{Consent for publication}
The authors consent to publish the paper.

\bmhead{Data availability}
\begin{itemize}
    \item Bill Smyth's String Repository - \url{https://www.cas.mcmaster.ca/~bill/strings}
    \item \textit{Pizza\&Chili} Text Collection - \url{https://pizzachili.dcc.uchile.cl/texts.html}
    \item Digits of $\pi$ - \url{https://calculat.io/storage/pi/100m.zip}
    \item Human Chromosome 1 - \url{https://www.ncbi.nlm.nih.gov/nuccore/NC_000001.11}
\end{itemize}
\bmhead{Code availability} \url{https://github.com/neerjamhaskar/closedStrings}
\
%
%





\bibliography{sn-bibliography}

\end{document}